\documentclass[twocolumn,showpacs,preprintnumbers,amsmath,amssymb,floatfix]{revtex4}

\usepackage[dvipdfmx]{graphicx}
\usepackage{bm}
\usepackage{color}
\usepackage{amsmath}
\usepackage{amsfonts}
\usepackage{amssymb}
\usepackage{ulem}

\usepackage[colorlinks=true,citecolor=blue,linkcolor=blue,urlcolor=blue,bookmarks=false]{hyperref}

\newcommand{\bvec}[1]{\mbox{\boldmath$#1$}} 
\def\scaleoffig{0.44}
\def\scaleoffigband{0.28285714285} 

\begin{document}

\preprint{APS/123-QED}

\title{Successive phase transitions and magnetization plateau in the spin-1 triangular-lattice antiferromagnet Ba$_2$La$_2$NiTe$_2$O$_{12}$ with small easy-axis anisotropy}

\author{Mutsuki Saito$^1$}
\email{saito@lee.phys.titech.ac.jp}
\author{Masari Watanabe$^1$}
\author{Nobuyuki Kurita$^1$}
\author{Akira Matsuo$^2$}
\author{Koichi Kindo$^2$}
\author{Maxim Avdeev$^{3,4}$}
\author{Harald O. Jeschke$^{5}$}
\author{Hidekazu Tanaka$^1$}
\email{tanaka@lee.phys.titech.ac.jp}

\affiliation{
$^1$Department of Physics, Tokyo Institute of Technology, Meguro-ku, Tokyo 152-8551, Japan\\
$^2$Institute for Solid State Physics, University of Tokyo, Kashiwa, Chiba 277-8581, Japan\\
$^3$Australian Nuclear Science and Technology Organisation, Lucas Heights, NSW 2234, Australia\\
$^4$School of Chemistry, The University of Sydney, Sydney 2006, Australia\\
$^5$Research Institute for Interdisciplinary Science, Okayama University, Kita-ku, Okayama 700-8530, Japan}
\date{\today}

\begin{abstract}
The crystal structure and magnetic properties of the spin-1 triangular-lattice antiferromagnet Ba$_2$La$_2$NiTe$_2$O$_{12}$ are reported. Its crystal structure is trigonal $R\bar{3}$, which is the same as that of Ba$_2$La$_2$NiW$_2$O$_{12}$ [Y. Doi {\it et al.}, \href{https://doi.org/10.1088/1361-648X/aa7c9b}{J. Phys.: Condens. Matter} \href{https://doi.org/10.1088/1361-648X/aa7c9b}{\textbf{29}}, \href{https://doi.org/10.1088/1361-648X/aa7c9b}{365802} (\href{https://doi.org/10.1088/1361-648X/aa7c9b}{2017})]. However, the exchange interaction $J/k_{\mathrm{B}}\simeq19$ K is much greater than that observed in the tungsten system. At zero magnetic field, Ba$_2$La$_2$NiTe$_2$O$_{12}$ undergoes successive magnetic phase transitions at $T_{\mathrm{N}1}=9.8$ K and $T_{\mathrm{N}2}=8.9$ K. The ground state is accompanied by a weak ferromagnetic moment. These results indicate that the ground-state spin structure is a triangular structure in a plane perpendicular to the triangular lattice owing to the small easy-axis-type anisotropy. The magnetization curve exhibits the one-third plateau characteristic of a two-dimensional triangular-lattice Heisenberg-like antiferromagnet. Exchange constants are also evaluated using density functional theory (DFT). The DFT results demonstrate the large difference in the exchange constants between tellurium and tungsten systems and the good two-dimensionality of the tellurium system.
\end{abstract}


\maketitle


\section{Introduction}
Triangular-lattice antiferromagnets (TLAFs) exhibit a variety of phase transitions in magnetic fields depending on magnetic anisotropy, spatial anisotropy and interlayer exchange interaction~\cite{Collins,Starykh2}. In particular, the magnetization plateau in TLAF has been attracting considerable attention. For two-dimensional (2D) classical spin TLAF with the easy-axis anisotropy, a magnetization plateau emerges at one-third of the saturation magnetization when a magnetic field is applied parallel to the easy axis~\cite{Miyashita}. The classical 1/3--magnetization plateau has been observed in quasi-2D large spin TLAFs GdPd$_2$Al$_3$~\cite{Kitazawa,Inami} and Rb$_4$Mn(MoO$_4$)$_3$~\cite{Ishii}. 

The easy-axis anisotropy is crucial for stabilizing the 1/3--magnetization plateau in the classical spin TLAF. The plateau is absent in the Heisenberg TLAF and Heisenberg-like TLAF with the easy-plane anisotropy. However, for 2D quantum spin Heisenberg TLAFs, the 1/3--magnetization plateau can be stabilized in a wide magnetic field range by quantum fluctuation~\cite{Nishimori,Chubokov,Nikuni,Honecker,Alicea,Farnell,Sakai,Richter,Hotta,Yamamoto1,Sellmann,Starykh2,Coletta}. The 1/3--magnetization plateau is affected by the magnetic anisotropy. When a magnetic field is applied parallel to the symmetry axis, the magnetic field range of the 1/3--magnetization plateau is enhanced by the easy-axis anisotropy and suppressed by the easy-plane anisotropy~\cite{Yamamoto1,Sellmann}. The quantum 1/3--magnetization plateau has actually been observed in quasi-2D spatially anisotropic TLAF Cs$_2$CuBr$_4$~\cite{Ono1,Ono2,Fortune} and uniform TLAF Ba$_3$CoSb$_2$O$_9$~\cite{Shirata,Zhou,Susuki,Quirion,Koutroulakis}, both of which have weak antiferromagnetic interlayer exchange interactions, and 3D TLAF CsCuCl$_3$~\cite{Sera} with strong ferromagnetic interlayer exchange interaction. All of these compounds have the weak easy-plane anisotropy.

Although the ground states in magnetic fields for the 2D spin-1/2 Heisenberg TLAF are well understood, the effects of the magnetic anisotropy~\cite{Yamamoto1,Sellmann}, spatial anisotropy~\cite{Ono2,Fortune,Starykh3}, interlayer exchange interaction~\cite{Susuki,Koutroulakis,Yamamoto2}, spin quantum number~\cite{Richter,Coletta} and thermal fluctuation on the ground states and phase diagram have not been sufficiently elucidated.

Recently, magnetic excitations in the spin-1/2 Heisenberg-like TLAF Ba$_3$CoSb$_2$O$_9$ were investigated by inelastic neutron scattering experiments~\cite{Zhou,Ma,Ito,Kamiya}. Unusual dynamical properties of single-magnon excitations predicted by theory such as the large downward quantum renormalization of excitation energies~\cite{Starykh,Zheng,Chernyshev,Mezio,Mourigal,Ghioldi} and a rotonlike minimum at the M point~\cite{Zheng,Ghioldi,Ghioldi2} were confirmed.
A notable feature of the magnetic excitations observed in Ba$_3$CoSb$_2$O$_9$ is a three-stage energy structure including intense dispersive excitation continua extending to a high energy six times the exchange constant~\cite{Ito}, which cannot be described by the current theory. These experimental results strongly indicate fractionalized spin excitations because the intense excitation continua cannot be explained in terms of conventional two-magnon excitations~\cite{Ghioldi2}. For the experimental elucidation of unconventional magnetic excitations, quantum TLAFs with different spin quantum numbers such as spin-1 are necessary. 

\begin{figure}[t!]
	\includegraphics[width=6.5cm]{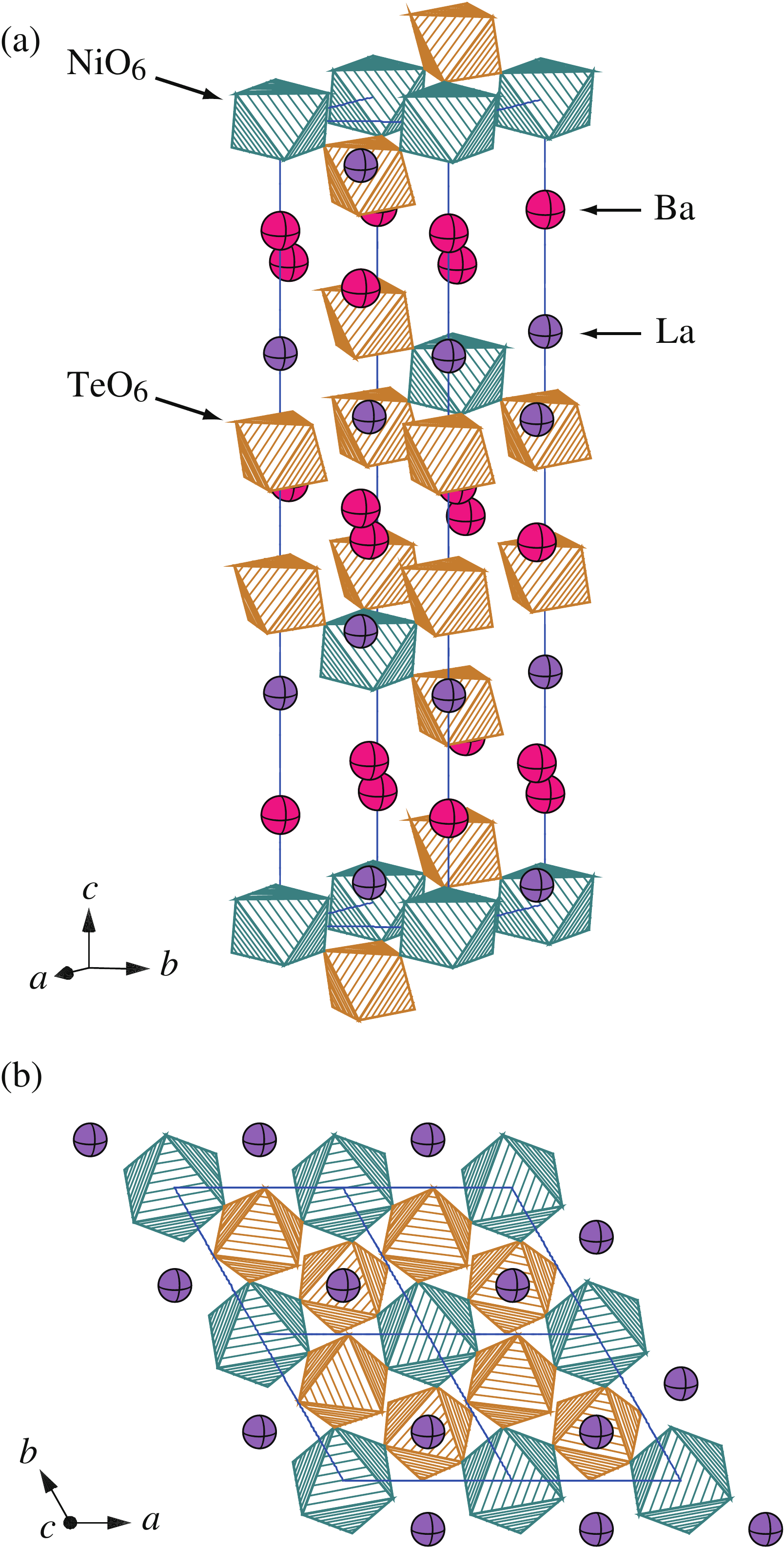}
	\caption{(Color online) (a) Schematic view of the crystal structure of Ba$_2$La$_2$NiTe$_2$O$_{12}$. The blue-green and ocher single octahedra are NiO$_6$ and TeO$_6$ octahedra with Ni$^{2+}$ and Te$^{6+}$ ions in the center, respectively.  Solid lines denote the chemical unit cell. (b) Crystal structure viewed along the $c$ axis. Magnetic Ni$^{2+}$ ions form a uniform triangular lattice in the $ab$ plane.}
	\label{fig:cryst}
\end{figure}

In this work, we investigated the crystal structure and magnetic properties of Ba$_2$La$_2$NiTe$_2$O$_{12}$. Although there is a brief report on the lattice constants and the space group of Ba$_2$La$_2$NiTe$_2$O$_{12}$~\cite{Autenrieth}, details of the crystal structure and magnetic properties have not been reported. The structure of this compound was found to be the same as that of Ba$_2$La$_2M$W$_2$O$_{12}$ ($M$\,=\,Mn, Co, Ni, Zn)~\cite{Sack,Li4,Rawl,Doi}, which have a uniform triangular lattice composed of transition metal ions $M^{2+}$. Figure \ref{fig:cryst} shows the crystal structure of Ba$_2$La$_2$NiTe$_2$O$_{12}$. An important feature of the crystal structure is that the magnetic triangular lattices are largely separated by layers of nonmagnetic ions; thus, we can expect good two-dimensionality. 

Recently, the magnetic properties in the family of triangular-lattice magnets Ba$_2$La$_2M$W$_2$O$_{12}$ ($M$\,=\,Mn, Co, Ni)~\cite{Rawl,Doi} have been investigated by magnetic susceptibility, specific heat and neutron diffraction (ND) measurements. Unfortunately, the exchange interactions were found to be weakly antiferromagnetic~\cite{Rawl} or weakly ferromagnetic~\cite{Doi}. It is natural to assume that superexchange interactions between neighboring spins in the same triangular layer occur through $M^{2+}-$\,O$^{2-}-$\,O$^{2-}-$\,$M^{2+}$ and $M^{2+}-$\,O$^{2-}-$\,W$^{6+}-$\,O$^{2-}-$\,$M^{2+}$ paths. The superexchange through the former path should be antiferromagnetic, while the latter path leads to a ferromagnetic superexchange interaction because the filled outermost orbitals of nonmagnetic W$^{6+}$ and Nb$^{5+}$ ions are $4p$ orbitals, as discussed in Refs.~\cite{Yokota,Koga}. It is considered that the superexchange interactions via these two paths almost cancel in the tungsten compounds, resulting in a weakly antiferromagnetic or ferromagnetic total exchange interaction. Meanwhile, when the nonmagnetic W$^{6+}$ ion is replaced by a Te$^{6+}$ ion, for which the filled outermost orbital is a $4d$ orbital, the superexchange interaction through the $M^{2+}-$\,O$^{2-}-$\,Te$^{6+}-$\,O$^{2-}-$\,$M^{2+}$ path becomes antiferromagnetic and the total exchange interaction should be strongly antiferromagnetic~\cite{Yokota,Koga}. 

This is our motivation for studying Ba$_2$La$_2$NiTe$_2$O$_{12}$. The exchange interaction in the triangular layer was found to be antiferromagnetic and strong as expected. We evaluated individual exchange constants using density functional theory (DFT). The DFT results demonstrate that the nearest-neighbor exchange interaction in the triangular layer is antiferromagnetic and predominant. As shown below, the 1/3--magnetization plateau characteristic of the quasi-2D TLAFs was observed in Ba$_2$La$_2$NiTe$_2$O$_{12}$. This compound is magnetically described as a quasi-2D spin-1 Heisenberg-like TLAF with small easy-axis-type anisotropy.

\section{Experimental details}

A powdered sample of Ba$_2$La$_2$NiTe$_2$O$_{12}$ was prepared by a solid-state reaction in accordance with the chemical reaction $2\mathrm{Ba}\mathrm{CO}_3+\mathrm{La}_2\mathrm{O}_3+\mathrm{Ni}\mathrm{O}+2\mathrm{Te}\mathrm{O}_2+\mathrm{O}_2\longrightarrow \mathrm{Ba}_2\mathrm{La}_2\mathrm{Ni}\mathrm{Te}_2\mathrm{O}_{12}+2\mathrm{CO}_2$ in air. $\mathrm{Ba}\mathrm{CO}_3$ (Wako, 99.9\%), $\mathrm{La}_2\mathrm{O}_3$ (Wako, 99.99\%), $\mathrm{Ni}\mathrm{O}$ (Wako, 99\%) and $\mathrm{Te}\mathrm{O}_2$ (Aldrich, 99.995\%) were mixed in stoichiometric quantities and calcined at 1000$^\circ$C in air for one day. Ba$_2$La$_2$NiTe$_2$O$_{12}$ was sintered at 1000$^{\circ}$C for one day after being pressed into a pellet. This sintering process was performed twice. Finally, yellow samples were obtained.

Powder X-ray diffraction (XRD) measurement of Ba$_2$La$_2$NiTe$_2$O$_{12}$ was conducted using a MiniFlex II diffractometer (Rigaku) with Cu $K\alpha$ radiation at room temperature. Powder ND measurement was also performed to determine both the crystal and magnetic structures using the high-resolution powder diffractometer Echidna installed at the OPAL reactor of the Australian Nuclear Science and Technology Organisation. The diffraction data were collected with a neutron wavelength of 2.4395 {\AA} in the temperature range of $1.6\,{\leq}\,T\,{\leq}\,14$ K. The crystal structure of Ba$_2$La$_2$NiTe$_2$O$_{12}$ was refined by Rietveld analysis of the powder XRD and ND data using the RIETAN-FP program~\cite{Izumi2007}.

Magnetic measurements in the temperature range of $1.8\,{\leq}\,T\,{\leq}\,300$ K and the magnetic field range of $0.1\,{\leq}\,\mu_0H\,{\leq}\,7.0$ T were performed using a Magnetic Property Measurement System (MPMS-XL, Quantum Design). High-field magnetization was measured  in a magnetic field of up to $\mu_0H=60$ T at $T\,{=}\,1.3$ K using an induction method with a multilayer pulse magnet at the Institute for Solid State Physics (ISSP), The University of Tokyo. Specific heat measurements in the temperature range of $1.9\,{\leq}\,T\,{\leq}\,300$ K  at magnetic fields of $\mu_0H\,{=}\,0$ and 9 T were performed using a Physical Property Measurement System (PPMS, Quantum Design) by the relaxation method.


\section{Computational details}

We determine the electronic structure of Ba$_2$La$_2$NiTe$_2$O$_{12}$ by performing all-electron DFT calculations based on the full potential local orbital (FPLO) code~\cite{Koepernik}. We use the generalized gradient approximation (GGA) exchange and correlation functional~\cite{Perdew}. The magnetic exchange interactions are determined by an energy-mapping method~\cite{Guterding2016,Iqbal2017,Iqbal2018}.
We account for the strong electronic correlations on the Ni $3d$ orbitals using the GGA+U exchange correlation functional~\cite{Liechtenstein} with the Hund's rule coupling strength $J_H=0.88$ eV fixed in accordance with the literature~\cite{Mizokawa}. The on-site interaction $U$ is determined using the experimental Curie--Weiss temperature as explained below. As the primitive rhombohedral unit cell of Ba$_2$La$_2$NiTe$_2$O$_{12}$ in the $R\,\bar{3}$ space group contains only a single Ni$^{2+}$ ion, we create supercells to allow spin configurations with different energies. A supercell containing four Ni$^{2+}$ ions provides four distinct energies and allows the resolution of nearest- and next-nearest-neighbor coupling in the triangular lattice. A supercell with six Ni$^{2+}$ ions and eight distinct energies is also required to resolve the shortest interlayer exchange path. As is common for triangular lattice antiferromagnets~\cite{Tapp2017}, the supercell calculations are computationally demanding, with each formula unit containing one magnetic ion adding more than 100 electrons to the calculation.

\section{Results and Discussion}

\subsection{Crystal structure}

\begin{figure}[t!]
	\centering
	\includegraphics[scale=\scaleoffig]{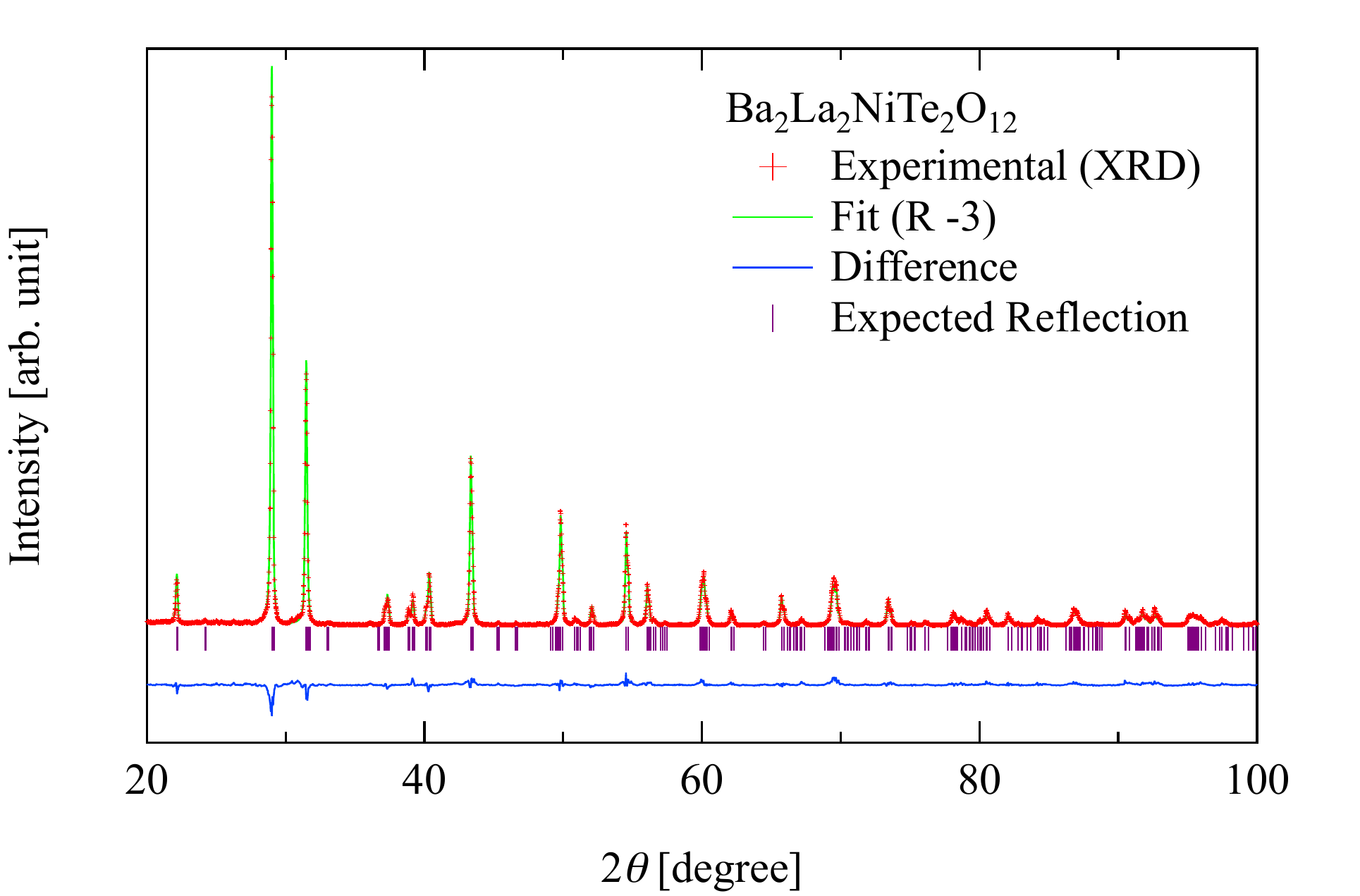}
	\caption{(Color online) XRD pattern of Ba$_2$La$_2$NiTe$_2$O$_{12}$ measured at room temperature. Experimental data, the results of Rietveld fitting, their difference and expected reflections are shown by the red symbols, green line, blue line and vertical purple bars, respectively.}
	\label{fig:XRD}
\end{figure}

\begin{table}[t!]
	\centering
	\caption{Structure parameters of $\mathrm{Ba}_2\mathrm{La}_2\mathrm{Ni}\mathrm{Te}_2\mathrm{O}_{12}$ determined from the XRD measurement at room temperature.}
	\label{tab:Rietveld_XRD}
	\begin{tabular}{lllll}
		\hline\hline
		Atom \ \ & Site & \ \ $x$ & \ \ $y$ & \ \ $z$ \\\hline
		$\mathrm{Ba}$\ \  & 6c\ \   & \ \ 0          &\ \  0        & \ \ 0.13587(7) \\
		$\mathrm{La}$ \ \ & 6c \ \  & \ \ 0          & \ \ 0        & \ \ 0.28973(6) \\
		$\mathrm{Ni}$ \ \ & 3a\ \   & \ \ 0          & \ \ 0        & \ \ 0          \\
		$\mathrm{Te}$ \ \ & 6c \ \  & \ \ 0          & \ \ 0        & \ \ 0.41560(7) \\
		$\mathrm{O(1)}$\ \  & 18f\ \  & \ \ 0.543(5) & \ \ 0.514(5) & \ \ 0.1186(3)  \\
		$\mathrm{O(2)}$\ \  & 18f\ \  & \ \ 0.450(5) & \ \ 0.473(5) & \ \ 0.2965(4) \\
		\hline 
		\multicolumn{5}{l}{\vspace{0.3mm}Space group $R{\bar 3}$}\\
		\multicolumn{5}{l}{$a=5.681(9)$ \AA, $c=27.60(3)$ \AA;}\\
		\multicolumn{5}{l}{$R_{\mathrm{wp}}=12.2\%$, $R_{\mathrm{p}}=9.4\%$, $R_{\mathrm{e}}=6.7\%$.}\\
		\multicolumn{5}{l}{$B=1.401$ \AA$^2$ for all atoms.}\\
		\hline\hline
	\end{tabular}
\end{table}

The results of the XRD measurement of Ba$_2$La$_2$NiTe$_2$O$_{12}$ at room temperature and the Rietveld analysis with RIETAN-FP~\cite{Izumi2007} are shown in Fig. \ref{fig:XRD}. First, we chose the structure parameters of Ba$_2$La$_2$NiW$_2$O$_{12}$~\cite{Rawl,Doi} as the initial parameters of the Rietveld analysis, setting the occupancy to 1 for all atoms and the thermal vibration parameter $B$ to $1.401$\,\AA$^2$, which was reported for Ba$_2$La$_2$NiW$_2$O$_{12}$~\cite{Rawl}. The analysis was based on two structural models with space groups $R\bar{3}m$ and $R\bar{3}$. It is difficult to determine the space group from only the XRD pattern because both structural models successfully reproduce the observed XRD pattern. However, the neutron diffraction pattern obtained at low temperatures above the first ordering temperature $T_{\mathrm{N}1}\,{\simeq}10$\,K is much better described by space group $R\bar{3}$ as shown below. 
The structure parameters refined for space group $R\bar{3}$ using the XRD data are summarized in Table \ref{tab:Rietveld_XRD}. 

\begin{figure}[t!]
	\centering
	\includegraphics[scale=\scaleoffig]{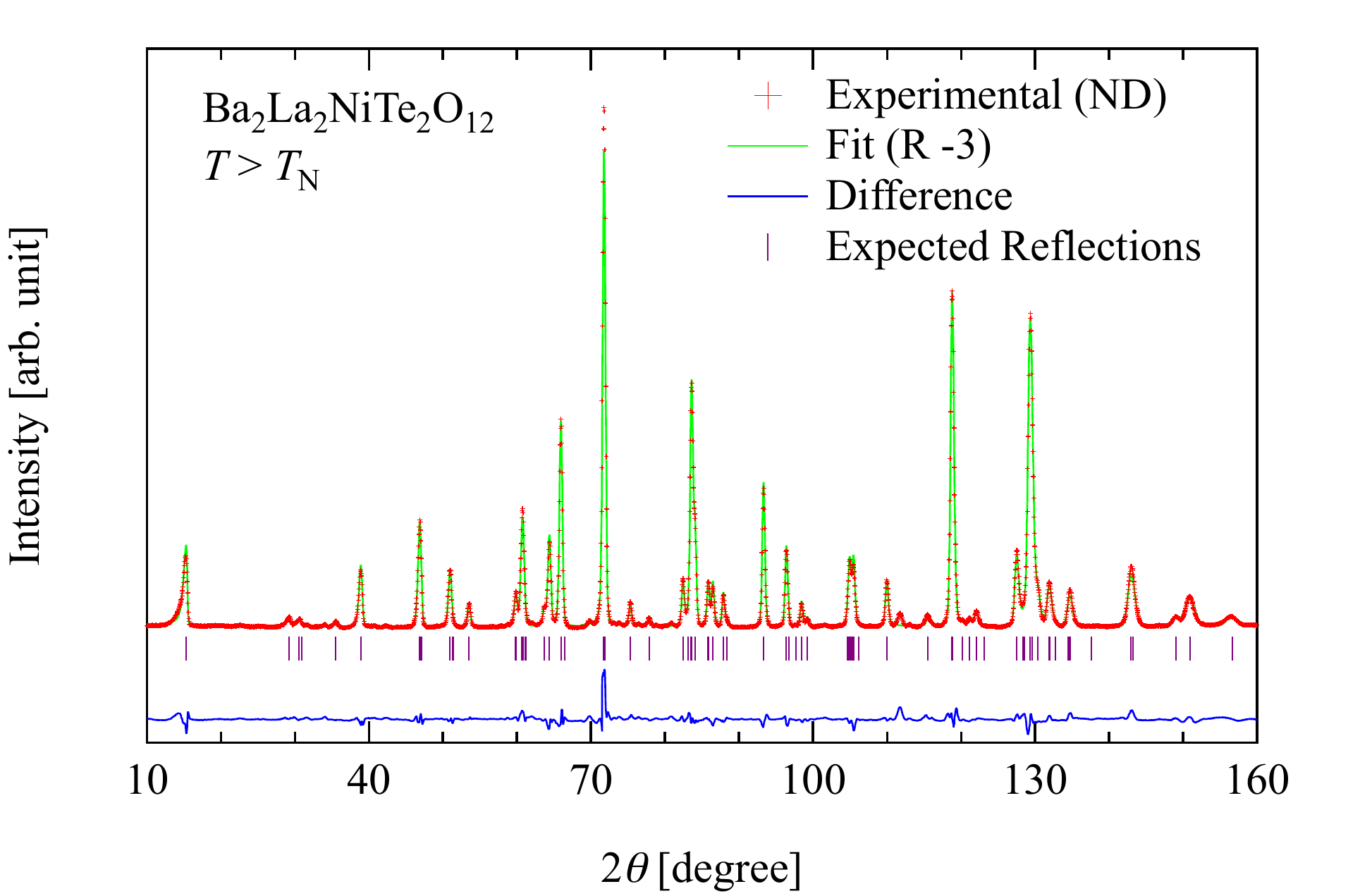}
	\caption{(Color online) ND pattern of Ba$_2$La$_2$NiTe$_2$O$_{12}$ measured at low temperatures above the first ordering temperature $T_{\mathrm{N}1}\,{=}\,9.8$ K. Experimental data, the results of Rietveld fitting, their difference and expected reflections are shown by the red symbols, green line, blue line and vertical purple bars, respectively. The experimental data is the average of measurements at $T\,{=}\,14, 12$ and 10 K.}
	\label{fig:ND}
\end{figure}

\begin{table}[t!]
	\centering
	\caption{Structure parameters of Ba$_2$La$_2$NiTe$_2$O$_{12}$ determined from the ND measurements at several temperatures above $T_{\mathrm{N}1}\,{\simeq}\,10$ K.}
	\label{tab:Rietveld_ND}
	\begin{tabular}{ccllll}
		\hline\hline
		Atom  & Site &  \ \ $x$ &  \ \ $y$ &  \ \ $z$ &  \ \ $B$ [\AA$^2$] \\\hline
		$\mathrm{Ba}$   \ \  &  6c  &  \ \ 0      &  \ \ 0      &  \ \ 0.1370(2) &  \ \ 0.354 \\
		$\mathrm{La}$   \ \  &  6c  &  \ \ 0       &  \ \ 0       &  \ \ 0.2890(1) &  \ \ 0.354 \\
		$\mathrm{Ni}$   \ \  & 3a  & \ \  0       & \ \  0       &  \ \ 0       &  \ \ 0.437 \\
		$\mathrm{Te}$   \ \  &  6c  & \ \  0       &  \ \ 0       &  \ \ 0.4150(1) &  \ \ 0.377 \\
		$\mathrm{O(1)}$  \ \ & 18f & \ \ 0.4631(4) & \ \ 0.4675(5) & \ \ 0.1168(1) & \ \ 0.877 \\
		$\mathrm{O(2)}$  \ \ & 18f & \ \ 0.4339(4) & \ \ 0.4603(5) & \ \ 0.2947(1) & \ \ 0.877 \\
		\hline 
		\multicolumn{5}{l}{\vspace{0.3mm}Space group $R{\bar 3}$}\\
		\multicolumn{5}{l}{$a=5.6682(7)$ \AA, $c=27.472(2)$ \AA;}\\
		\multicolumn{5}{l}{$R_{\mathrm{wp}}=7.9\%$, $R_{\mathrm{p}}=5.7\%$, $R_{\mathrm{e}}=1.5\%$.}\\
		\hline\hline
	\end{tabular}
\end{table}

Figure \ref{fig:ND} shows the ND pattern of Ba$_2$La$_2$NiTe$_2$O$_{12}$ measured at low temperatures above the first ordering temperature $T_{\mathrm{N}1}\,{=}9.8$ K, where the diffraction intensity is the average of those measured at $T\,{=}\,14, 12$ and 10 K. We analyzed the ND data on the basis of two structural models with space groups $R\bar{3}m$ and $R\bar{3}$. The values of $R_{\mathrm{wp}}$ and $R_{\mathrm{p}}$ are obtained from the refinements to be 22.1\% and 15.3\% for $R\bar{3}m$ and 7.9\% and 5.7\% for $R\bar{3}$, respectively. The $R$-factors for $R\bar{3}$ are significantly smaller than those for $R\bar{3}m$. Because no structural phase transition was detected via magnetic susceptibility and specific heat measurements down to 1.8 K, we can conclude that the space group of Ba$_2$La$_2$NiTe$_2$O$_{12}$ is $R\bar{3}$, which is the same as the space group of Ba$_2$La$_2$MW$_2$O$_{12}$ (M=Mn, Co, Ni, Zn)~\cite{Doi}.
The difference between the crystal structures for these space groups is in the atomic positions of oxygen atoms. Because the atomic scattering factor of oxygen atoms for X-rays is much smaller than those of other atoms, it is difficult to determine the atomic positions of oxygen accurately by XRD measurement, as pointed out by Doi \textit{et al.}~\cite{Doi}. For $R\bar{3}$, NiO$_6$ and TeO$_6$ octahedra are rotated in opposite directions around the $c$ axis, which leads to the absence of mirror symmetry, as shown in Fig. \ref{fig:cryst}(b). The structure parameters refined for space group $R\bar{3}$ using the ND data are summarized in Table \ref{tab:Rietveld_ND}.


\subsection{Magnetic susceptibility and low-field magnetization}

\begin{figure}[t!]
	\centering
	\includegraphics[scale=\scaleoffig]{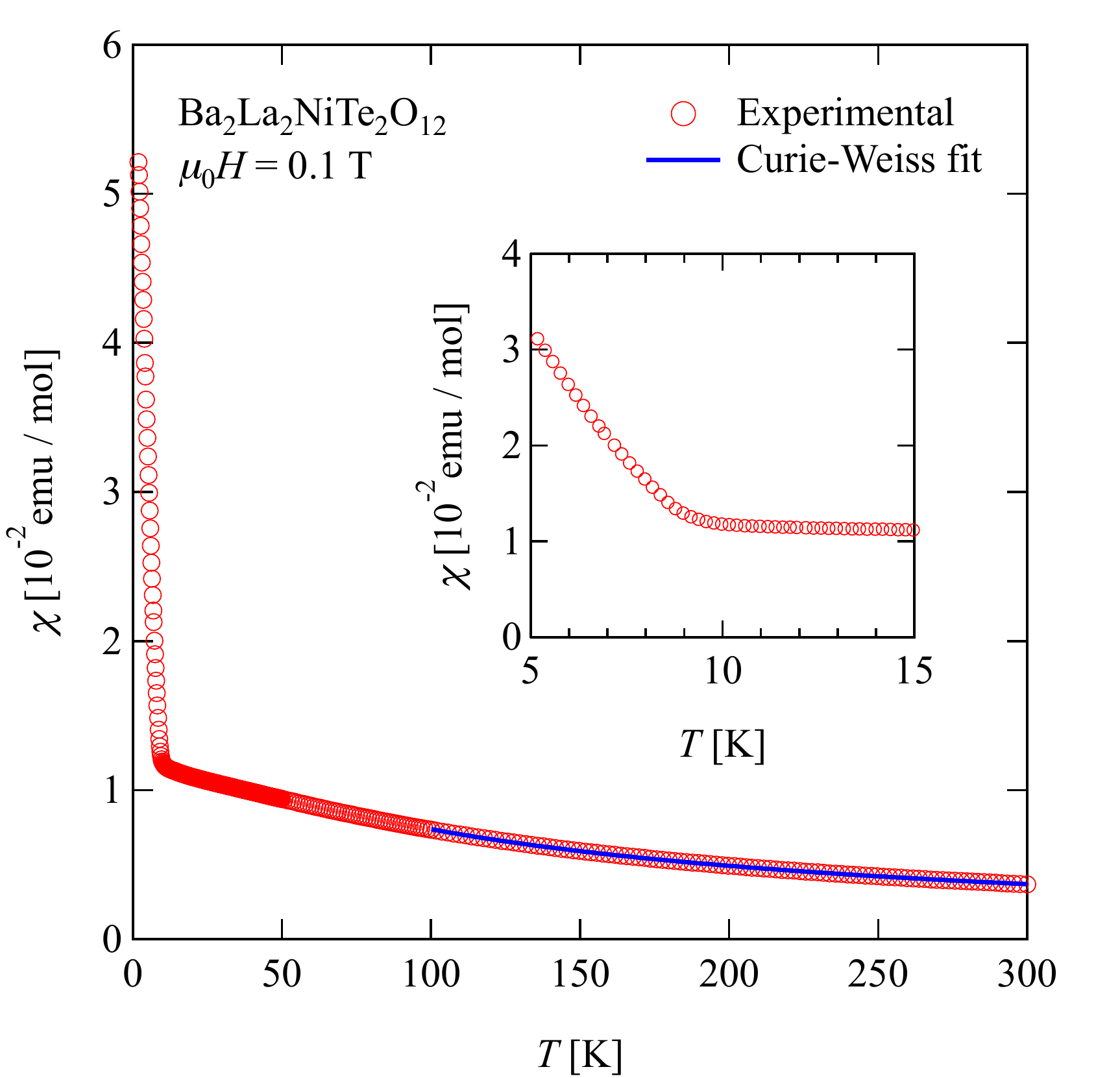}
	\caption{(Color online) Temperature dependence of the magnetic susceptibility of Ba$_2$La$_2$NiTe$_2$O$_{12}$ powder measured in an external magnetic field of $\mu_0H\,{=}\,0.1$ T. The blue solid line shows the result of a Curie--Weiss fit in the temperature range of $100\,{\leq}\,T\,{\leq}\,300$ K. The inset is an enlarged view around 10 K.}
	\label{fig:MT}
\end{figure}

The temperature dependence of the magnetic susceptibility of Ba$_2$La$_2$NiTe$_2$O$_{12}$ powder measured in a magnetic field of $\mu_0H\,{=}\,0.1$ T is shown in Fig. \ref{fig:MT}. The Curie constant $C\,{=}\,1.482(2)$ emu K mol$^{-1}$ and the Weiss temperature $\Theta_{\mathrm{CW}}\,{=}\,-100.7(3)$ K were obtained by fitting to the Curie--Weiss law $\chi(T)\,{=}\,C/(T-\Theta_{\mathrm{CW}})$ in the temperature range $100\,{\leq}\,T\,{\leq}\,300\,{\mathrm{K}}$. This large negative $\Theta_{\mathrm{CW}}$ indicates that the dominant exchange interaction of Ba$_2$La$_2$NiTe$_2$O$_{12}$ is antiferromagnetic and large, as expected from the superexchange path via the filled outermost $4d$ orbital of Te$^{6+}$. The exchange constant $J$, effective magnetic moment $\mu_{\mathrm{eff}}$ and $g$-factor are estimated as $J/k_{\mathrm{B}}\,{=}\,25$ K, $\mu_{\mathrm{eff}}\,{=}\,3.44\,\mu_{\mathrm{B}}$ and $g\,{=}\,2.4$ on the basis of molecular field theory. 

The magnetic susceptibility of Ba$_2$La$_2$NiTe$_2$O$_{12}$ increases rapidly near 9 K as the temperature decreases, which is indicative of the antiferromagnetic phase transition. This transition temperature of $T_{\mathrm{N}}\,{\simeq}\,9$ K is lower than $T_{\mathrm{N}}\,{\simeq}\,13$ K for Ba$_3$NiSb$_2$O$_9$ ~\cite{Shirata2,Doi2}, which is an $S\,{=}\,1$ TLAF with a crystal structure and exchange interaction $J/k_{\mathrm{B}}\,{\simeq}\,20$ K, similar to those of Ba$_2$La$_2$NiTe$_2$O$_{12}$~\cite{Shirata2,Doi2,Richter}. Thus, the two-dimensionality in Ba$_2$La$_2$NiTe$_2$O$_{12}$ is better than that in Ba$_3$NiSb$_2$O$_9$. Note that the magnetic susceptibility of Ba$_3$NiSb$_2$O$_9$ powder does not show a rapid upturn below $T_{\mathrm{N}}$~\cite{Doi2}.

\begin{figure}[t!]
	\centering
	\includegraphics[width=8.0cm]{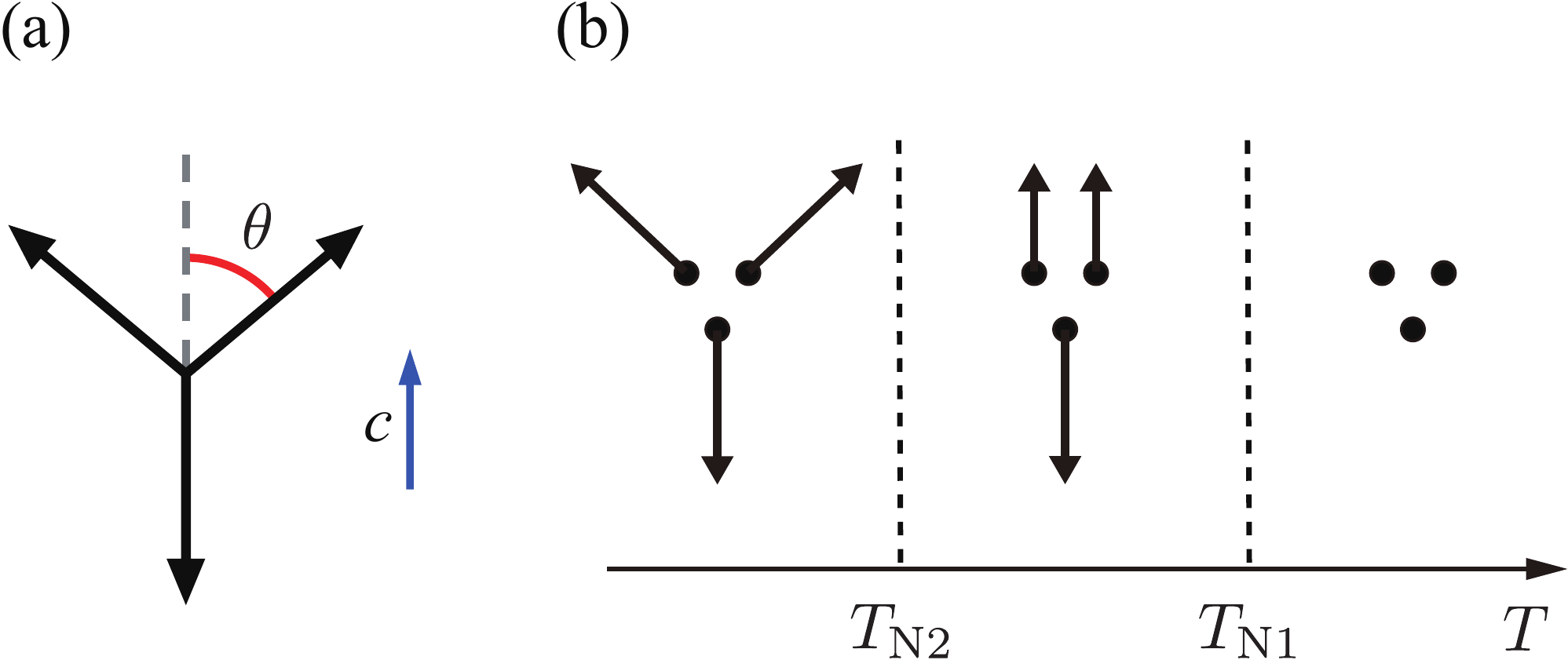}
	\caption{(Color online) (a) Triangular structure in a plane including the $c$ axis for TLAF with small easy-axis-type anisotropy. The angle ${\theta}$ between canted sublattice spins and the $c$ axis is smaller than $60^{\circ}$. (b) Schematic view of the successive magnetic phase transitions in the TLAF with small easy-axis-type anisotropy and the spin structures in each phase.}
	\label{fig:EasyAxis}
\end{figure}

A notable feature of the magnetic susceptibility in Ba$_2$La$_2$NiTe$_2$O$_{12}$ is the rapid increase below $T_{\mathrm{N}}$. This behavior can be understood in terms of a small easy-axis-type anisotropy and a ferromagnetic interlayer exchange interaction. 
When the magnetic anisotropy is of the easy-axis type and small, the spin configuration in the ground state is a triangular structure in a plane including the crystallographic $c$ axis, as shown in Fig. \ref{fig:EasyAxis}(a). The triangular structure is slightly distorted from a perfect $120^\circ$ structure. The angle ${\theta}$ between canted sublattice spins and the $c$ axis is smaller than $60^\circ$. Therefore, the sum of the magnetic moments of three sublattice spins is nonzero; thus, a resultant magnetic moment along the $c$ axis appears in a triangular layer. When the interlayer exchange interaction is antiferromagnetic, the resultant magnetic moments appearing in the neighboring triangular layers are canceled out. On the other hand, when the interlayer exchange interaction is ferromagnetic, all the resultant magnetic moments appearing in the triangular layers align in the same direction, giving the system a net magnetic moment along the $c$ axis. The small easy-axis-type anisotropy of Ba$_2$La$_2$NiTe$_2$O$_{12}$ is also consistent with the successive magnetic phase transitions observed by the specific heat measurements shown later.


\begin{figure}[t!]
	\centering
	\includegraphics[scale=\scaleoffig]{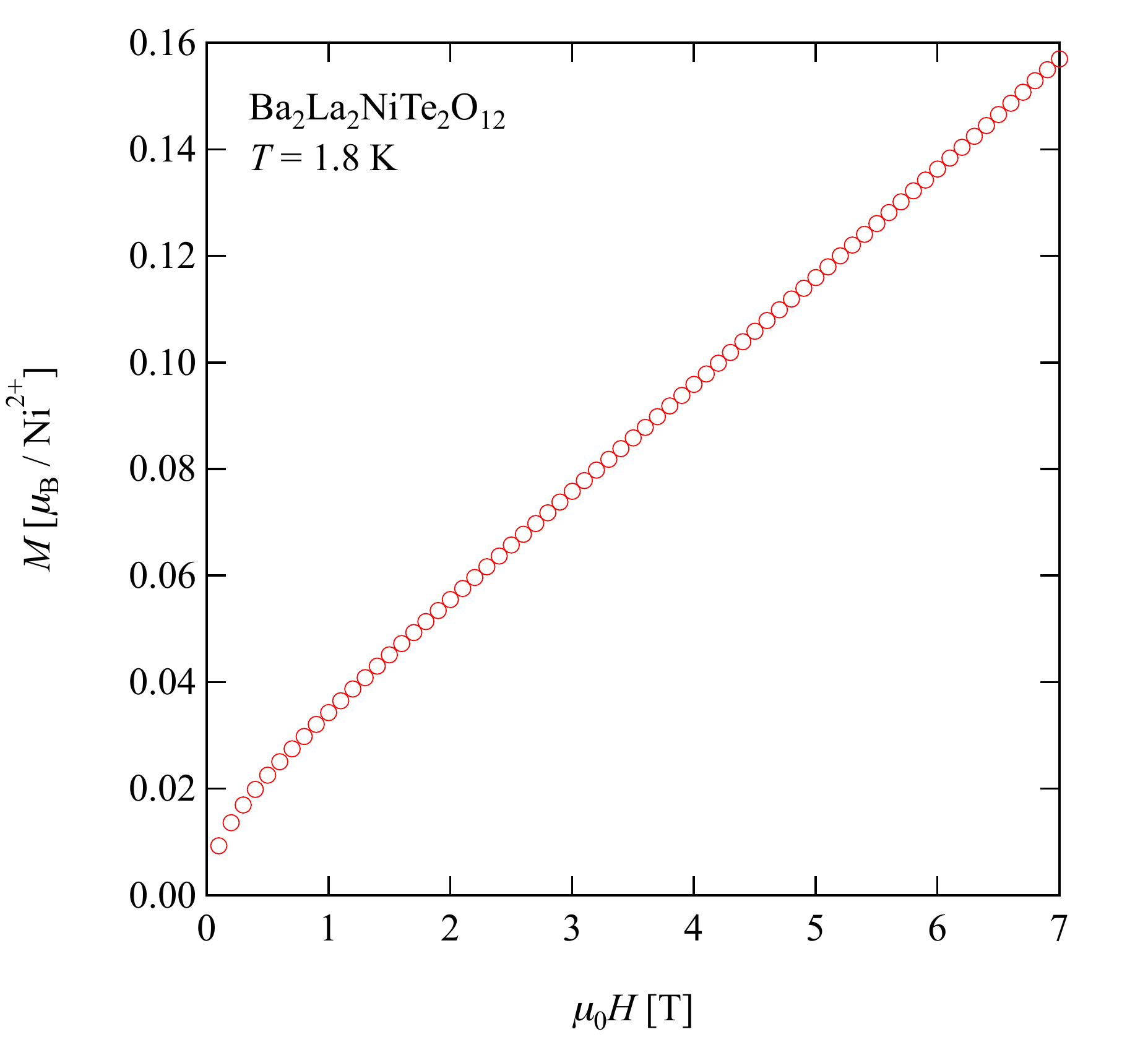}
	\caption{(Color online) Magnetization curve of Ba$_2$La$_2$NiTe$_2$O$_{12}$ powder measured at $T\,{=}\,1.8$ K in magnetic fields up to $\mu_0H\,{=}\,7$ T.}
	\label{fig:MH}
\end{figure}

The magnetic field dependence of the magnetization of Ba$_2$La$_2$NiTe$_2$O$_{12}$ powder is shown in Fig. \ref{fig:MH}. It is clearly observed that there is a finite magnetization even in zero field. The magnetic moment per spin $\Delta M$ in the ground state at zero magnetic field is given by
\begin{equation}
	\Delta M = \frac{1}{3}(2\cos \theta-1) g\mu_{\mathrm{B}} S,
\end{equation}
where $S\,{=}\,1$ and $\theta$ is the canting angle shown in Fig. \ref{fig:EasyAxis}(a). The powder average of the weak moment $\overline{\Delta M}$ is given by $\overline{\Delta M}\,{=}\,\Delta M/2$. By using the value $\overline{\Delta M}\,{=}\,0.015\,{\mu_{\mathrm{B}}}/{\mathrm{Ni}}^{2+}$, which is obtained by extrapolating the magnetization curve to zero magnetic field, and $g\,{=}\,2.4$ estimated from the Curie constant, we obtain the angle ${\theta}\,{=}\,58.75^\circ$.


The origin of the small easy-axis-type anisotropy is considered to be the single-ion anisotropy expressed as $D\left(S_i^z\right)^2$ with $D\,{<}\,0$.
The canting angle $\theta$ is expressed as
\begin{equation}
	\cos\theta=\frac{3J}{6J-2|D|}.
\end{equation}
Using ${\theta}\,{=}\,58.75^\circ$, we obtain $|D|/J\,{=}\,0.108$.

\subsection{Specific heat}

\begin{figure}[t!]
	\centering
	\includegraphics[scale=\scaleoffig]{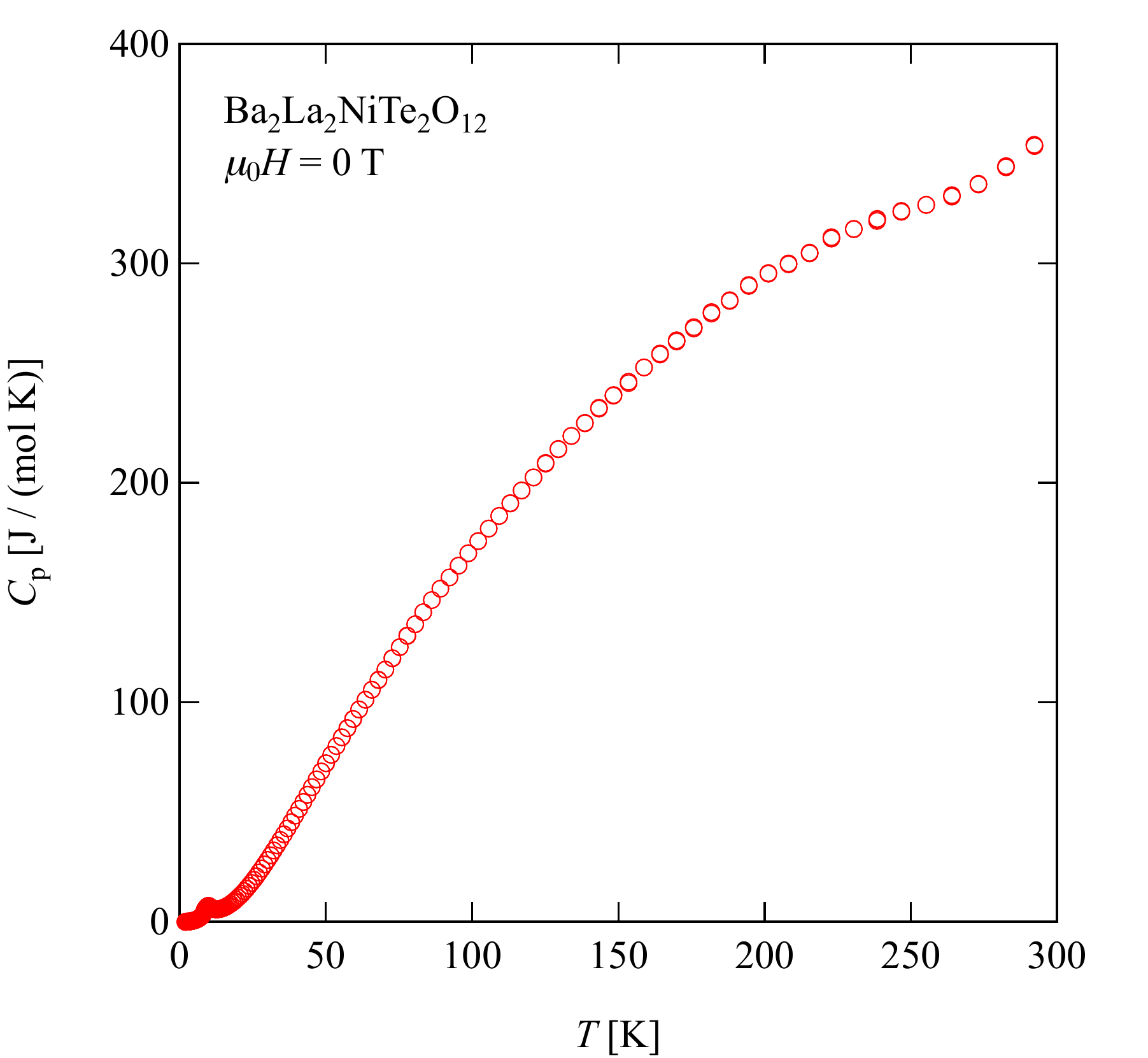}
	\caption{(Color online) Temperature dependence of the total specific heat of Ba$_2$La$_2$NiTe$_2$O$_{12}$ powder below 300 K measured at zero magnetic field.}
	\label{fig:HC}
\end{figure}

\begin{figure}[t!]
	\centering
	\includegraphics[scale=\scaleoffig]{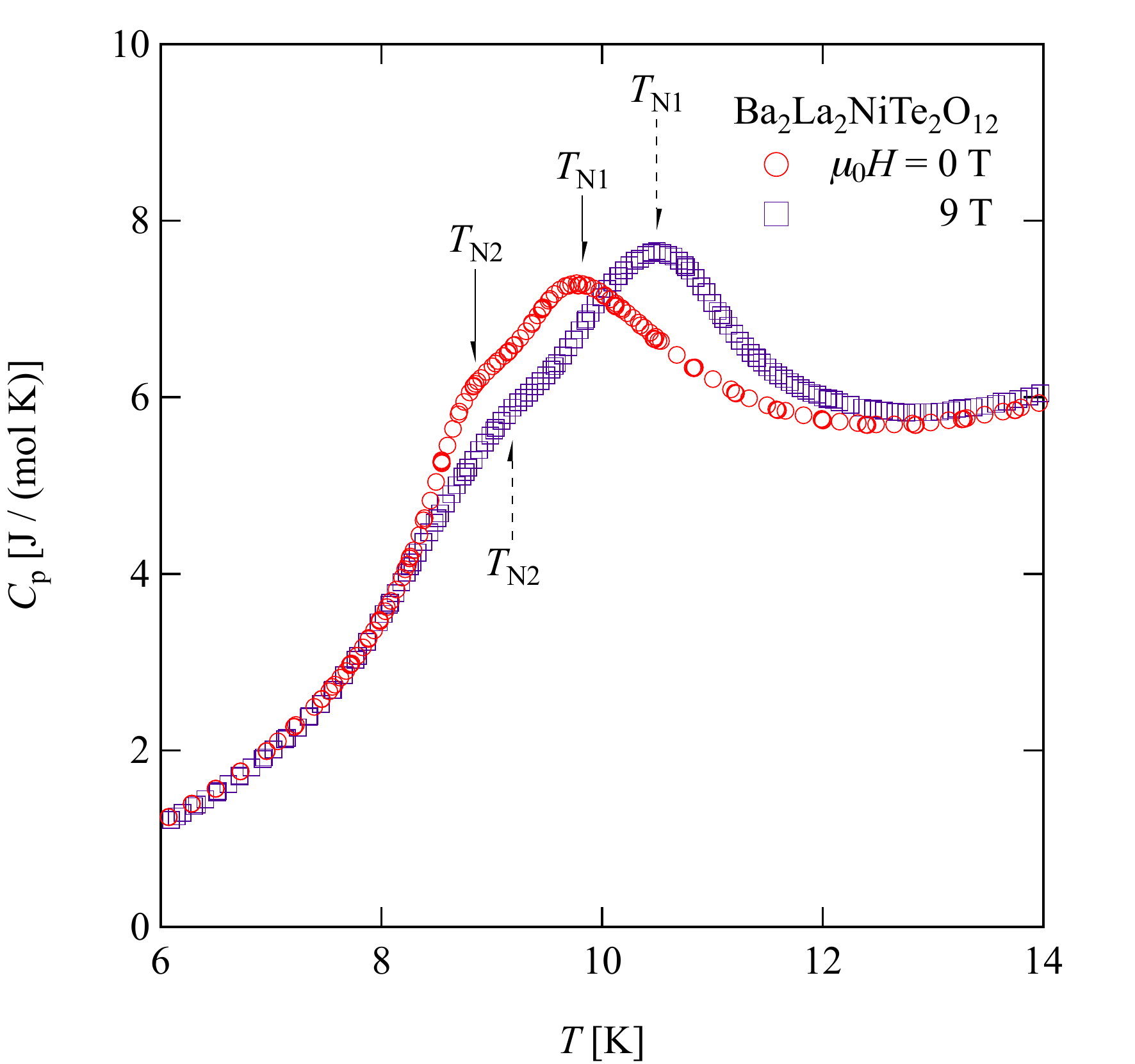}
	\caption{(Color online) Low-temperature specific heat of Ba$_2$La$_2$NiTe$_2$O$_{12}$ powder measured at $\mu_0H\,{=}\,0$ and 9 T. Arrows indicate magnetic phase transition temperatures $T_{\mathrm{N}1}$ and $T_{\mathrm{N}2}$.}
	\label{fig:HC_zoom}
\end{figure}

The temperature dependence of the specific heat of Ba$_2$La$_2$NiTe$_2$O$_{12}$ powder below 300 K measured at zero magnetic field is shown in Fig. \ref{fig:HC}. There is no anomaly indicative of a structural phase transition below 300 K. The hump anomaly around room temperature is an extrinsic anomaly that originates from the instability of the temperature. The low-temperature specific heat measured at $\mu_0H\,{=}\,0$ and 9 T is shown in Fig. \ref{fig:HC_zoom}. Double peaks indicative of successive magnetic phase transitions are observed at $T_{\mathrm{N}1}\,{=}\,9.8$ K, $T_{\mathrm{N}2}\,{=}\,8.9$ K for $\mu_0H\,{=}\,0$ T and at $T_{\mathrm{N}1}\,{=}\,10.5$ K, $T_{\mathrm{N}2}\,{=}\,9.2$ K for $\mu_0H\,{=}\,9$ T. Each transition temperature shifts to the high-temperature side with increasing magnetic field, and the shift for $T_{\mathrm{N}1}$ is larger than that for $T_{\mathrm{N}2}$. 

It is theoretically known that successive magnetic phase transitions occur in a TLAF with easy-axis-type anisotropy~\cite{Miyashita,Matsubara}. With decreasing temperature, the $z$ components of spins order first at $T\,{=}\,T_{\mathrm{N}1}$, and the $xy$ components of spins order next at $T\,{=}\,T_{\mathrm{N}2}$, as shown in Fig. \ref{fig:EasyAxis}(b).
Similar successive magnetic phase transitions arising from the small easy-axis-type anisotropy were reported for Ba$_3$NiSb$_2$O$_9$~\cite{Shirata2}, which has an exchange constant similar to that of Ba$_2$La$_2$NiTe$_2$O$_{12}$~\cite{Shirata2,Richter}. The phase transition temperatures of Ba$_3$NiSb$_2$O$_9$ are $T_{\mathrm{N}1}\,{=}\,13.5$ K and $T_{\mathrm{N}2}\,{=}\,13.0$ K, both of which are higher than those of Ba$_2$La$_2$NiTe$_2$O$_{12}$. This suggests that the two-dimensionality in Ba$_2$La$_2$NiTe$_2$O$_{12}$ is better than that in Ba$_3$NiSb$_2$O$_9$.

Using molecular field theory~\cite{Matsubara}, two transition temperatures are calculated as $T_{\rm N1}\,{=}\,38.7$ K and $T_{\rm N2}\,{=}\,37.6$ K with $|D|/J\,{=}\,0.108$ and the saturation field $H_{\rm s}\,{=}\,110$ T obtained below. Although their absolute values are four times larger than those observed, their separation of $T_{\rm N1}\,{-}\,T_{\rm N2}\,{=}\,1.1$ K is consistent with the experimental separation of 0.9 K.

\begin{figure}[t!]
	\centering
	\includegraphics[scale=\scaleoffig]{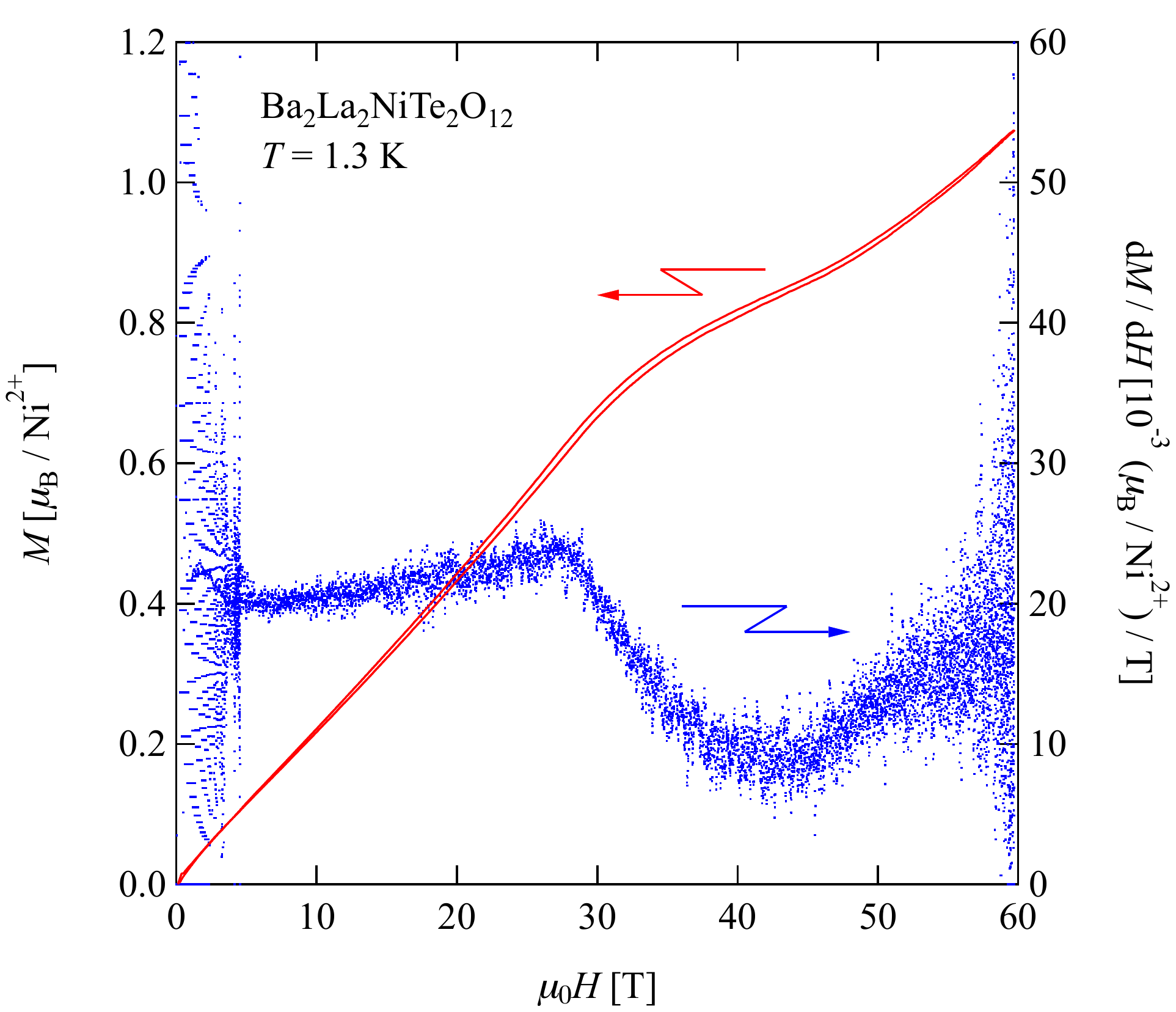}
	\caption{(Color online) High-field magnetization process of Ba$_2$La$_2$NiTe$_2$O$_{12}$ measured at 1.3 K upon sweeping the magnetic field up and down. Red solid lines and blue points are the magnetization $M$ and its field derivative $\mathrm{d}M/\mathrm{d}H$, respectively.}
	\label{fig:HighField}
\end{figure}

\subsection{High-field magnetization}

The result of the high-field magnetization measurement of Ba$_2$La$_2$NiTe$_2$O$_{12}$ powder up to 60 T is shown in Fig.~\ref{fig:HighField}. The absolute value of the magnetization is calibrated by using the result of the magnetization measurement up to 7 T with a SQUID magnetometer. A magnetization plateau is clearly observed at $M\,{\simeq}\,0.8$\,$\mu_{\mathrm{B}}$/Ni$^{2+}$ for $32\,{<}\,\mu_0H\,{<}\,47$ T. The lower and higher edge fields of the plateau were assigned to the magnetic fields at which $\mathrm{d}M/\mathrm{d}H$ has inflection points. Because the $g$-factor estimated from the magnetic susceptibility is $g\,{=}\,2.4$, the plateau corresponds to the 1/3--magnetization plateau characteristic of the quasi-2D TLAF. The edge fields of the plateau are rather smeared and the plateau is not completely flat. It is expected that this arises from the distribution of the edge fields in the powdered sample owing to the anisotropy of the $g$-factors and the magnetic anisotropy and not from exchange randomness~\cite{Watanabe,Kawamura}. When the anisotropy of the $g$-factor is $\Delta g$, the edge fields $H_{\mathrm{c}\alpha}$ with ${\alpha}\,{=}\,1$ and 2 are distributed in the range of $(\Delta g/\bar{g})H_{\mathrm{c}\alpha}$, where $\bar{g}$ is the average of the $g$-factor. When the magnetic anisotropy is of the easy-axis type, the field range of the 1/3--plateau becomes wider for $H\,{\parallel}\,c$ and narrower for $H\,{\perp}\,c$ when compared to the Heisenberg model.

Although the classical Heisenberg-like TLAF with easy-axis anisotropy exhibits the 1/3--magnetization plateau, it is difficult to explain the observed magnetization process in terms of a classical spin model only~\cite{Miyashita}. The lower and higher edge fields of the classical plateau are calculated as ${\mu_0}H_{\rm c1}\,{=}\,34.9$ T and ${\mu_0}H_{\rm c2}\,{=}\,44.6$ T with $|D|/J\,{=}\,0.108$ and the saturation field ${\mu_0}H_{\rm s}\,{=}\,110$ T obtained below. The width of the classical plateau is estimated as ${\mu_0}(H_{\rm c2}\,{-}\,H_{\rm c1})\,{=}\,9.7$ T, which is 65 \% of observed width of 15 T. It is known that at finite temperature, thermal fluctuation stabilizes the UUD spin state even in the classical spin model, so that the field range of the UUD state increases with increasing temperature~\cite{Kawamura2,Seabra}. However, in the present case, the effect of the thermal fluctuation should be negligible because the temperature of the magnetization measurement $T\,{=}\,1.3$ K is much lower than $T_{\rm N2}\,{=}8.9$ K.

\begin{figure}[t!]
	\centering
	\includegraphics[scale=\scaleoffig]{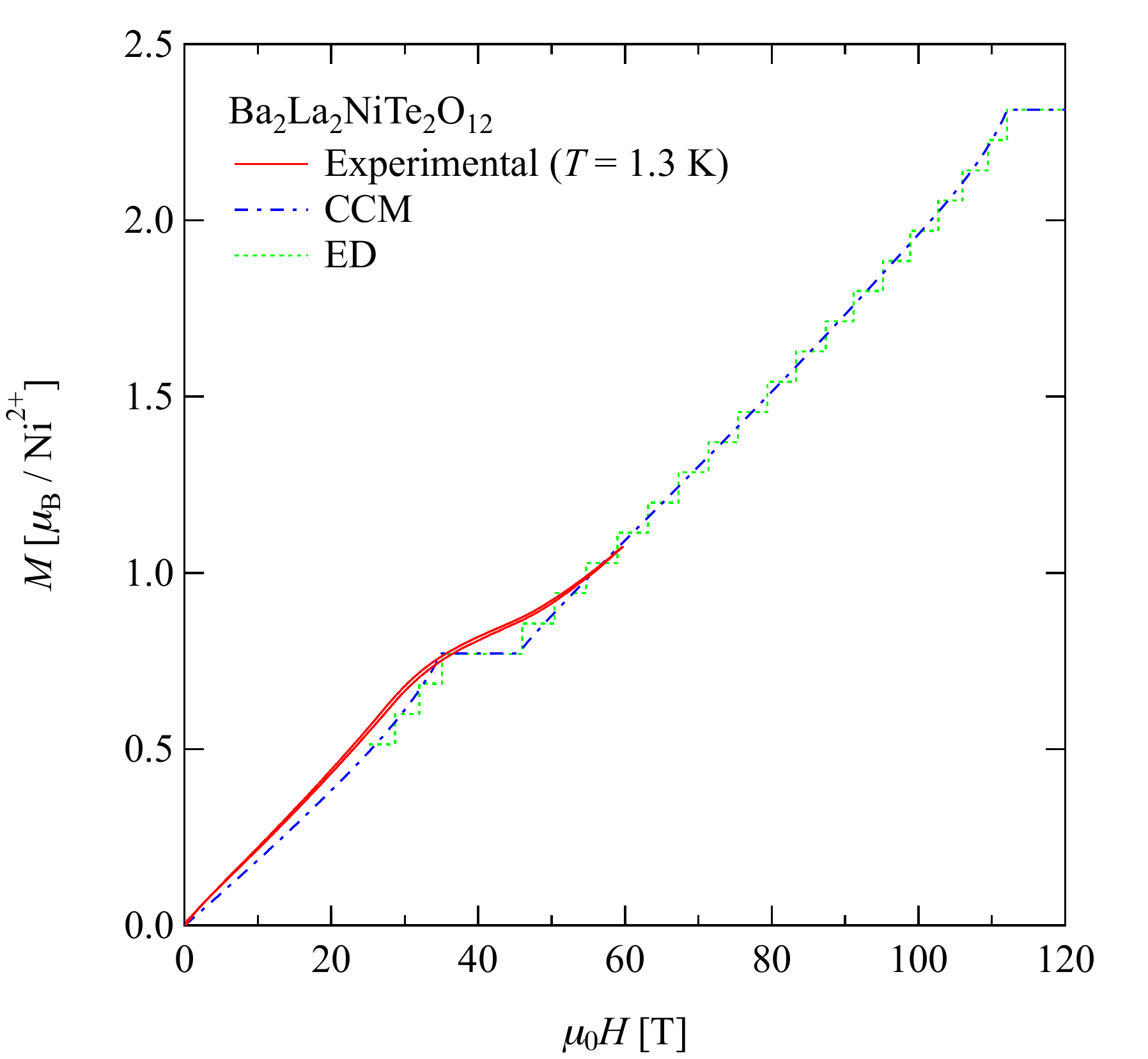}
	\caption{(Color online) Experimental magnetization curve of Ba$_2$La$_2$NiTe$_2$O$_{12}$ powder up to 60 T (red line) and theoretical magnetization curves of the $S\,{=}\,1$ Heisenberg TLAF calculated by CCM (blue line) and ED (green line)~\cite{Richter}.}
	\label{fig:HighField_fit}
\end{figure}

For a spin-1 Heisenberg TLAF, the 1/3--magnetization plateau is stabilized in a fairly wide magnetic field range by quantum fluctuations~\cite{Richter,Coletta}. We fit the theoretical magnetization curves of the spin-1 Heisenberg TLAF calculated by the coupled cluster method (CCM) and the exact diagonalization (ED)~\cite{Richter} to our experimental result, as shown in Fig.~\ref{fig:HighField_fit}. From this fit, we obtain ${\mu_0}H_{\rm c1}\,{=}\,35$ T, ${\mu_0}H_{\rm c2}\,{=}\,46$ T, ${\mu_0}H_{\rm s}\,{=}\,110$ T and the saturation magnetization $M_{\mathrm{s}}/\mu_{\mathrm{B}}\,{=}\,2.31(2)$, which leads to $g\,{=}\,2.31(2)$. The saturation magnetic field $H_{\mathrm{s}}$ of the spin-1 Heisenberg TLAF is given by $g\mu_{\mathrm{B}} H_{\mathrm{s}}\,{=}\,9JS$. Using $g\,{\simeq}\,2.3$ and $H_{\mathrm{s}}\,{\simeq}\,110$ T, which are estimated from the theoretical magnetization curve fitted to the magnetization data, the exchange interaction is estimated as $J/k_{\mathrm{B}}\,{\simeq}\,19$ K. This $J$ value is somewhat smaller than $J/k_{\mathrm{B}}\,{=}\,25$ K estimated from the Weiss constant ${\Theta_{\mathrm{CW}}}\,{=}\,{-}\,100.7$ K of the high-temperature magnetic susceptibility. Because the saturation field given by $g\mu_{\mathrm{B}} H_{\mathrm{s}}\,{=}\,9JS$ is exact, the exchange constant $J/k_{\mathrm{B}}\,{\simeq}\,19$ K estimated from the saturation field is considered to be more precise.

The magnetic field range of the experimental 1/3--plateau $32\,{<}\,\mu_0H\,{<}\,47$ T is somewhat larger than the field ranges $35\,{<}\,{\mu_0}H\,{<}\,46$ T and $34.9\,{<}\,{\mu_0}H\,{<}\,44.6$ T calculated on the basis of the spin-1 Heisenberg TLAF and the classical Heisenberg-like TLAF with $|D|/J\,{=}\,0.108$, respectively. Recent theory demonstrates that when a magnetic field is applied parallel to the symmetry axis, the field range of the quantum 1/3--magnetization plateau is enhanced by the easy-axis anisotropy and suppressed by the easy-plane anisotropy~\cite{Yamamoto1,Sellmann}. Thus, it is suggested that the synergy between quantum fluctuation and the easy-axis anisotropy makes the field range of the 1/3--plateau wider for $H\,{\parallel}\,c$ in Ba$_2$La$_2$NiTe$_2$O$_{12}$. On the other hand, the easy-axis anisotropy will act to suppress the plateau width for $H\,{\perp}\,c$. Thus, it is considered that the plateau width depends on the angle between the magnetic field and the $c$ axis, which leads to the distribution of the lower and higher edge fields $H_{\mathrm{c1}}$ and $H_{\mathrm{c2}}$ in a powdered sample. In addition, in case that the magnetic field is not exactly parallel to the $c$ axis, the total spin is not conserved. Consequently, the 1/3--plateau does not become completely flat and has finite slope. These factors will give rise to the smearing of the 1/3--plateau in a powdered sample, as observed in the present measurement.

\subsection{Magnetic structure}

\begin{figure}[t!]
	\centering
	\includegraphics[scale=\scaleoffig]{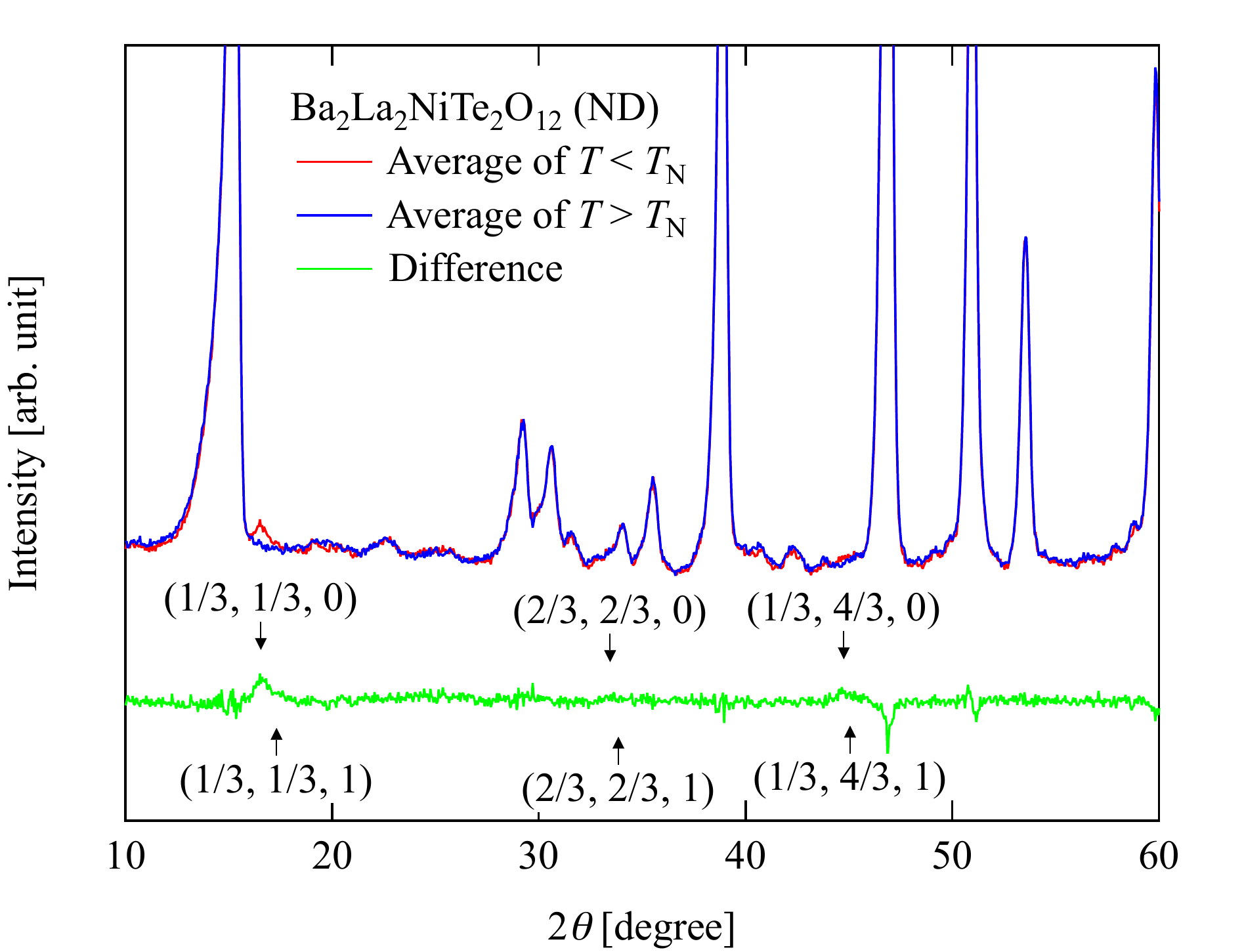}
	\caption{(Color online) ND intensities of Ba$_2$La$_2$NiTe$_2$O$_{12}$ powder averaged over $T\,{=}\,6, 4$ and 1.6 K $(<T_{\mathrm{N}2})$ (red) and $T\,{=}\,14, 12$ and 10 K $(>T_{\mathrm{N}1})$ (blue). Their difference is drawn by the green line. Arrows denote the positions of magnetic Bragg peaks with the indicated wave vectors.}
	\label{fig:ND_TN}
\end{figure}

\begin{figure}[t!]
	\centering
	\includegraphics[scale=\scaleoffig]{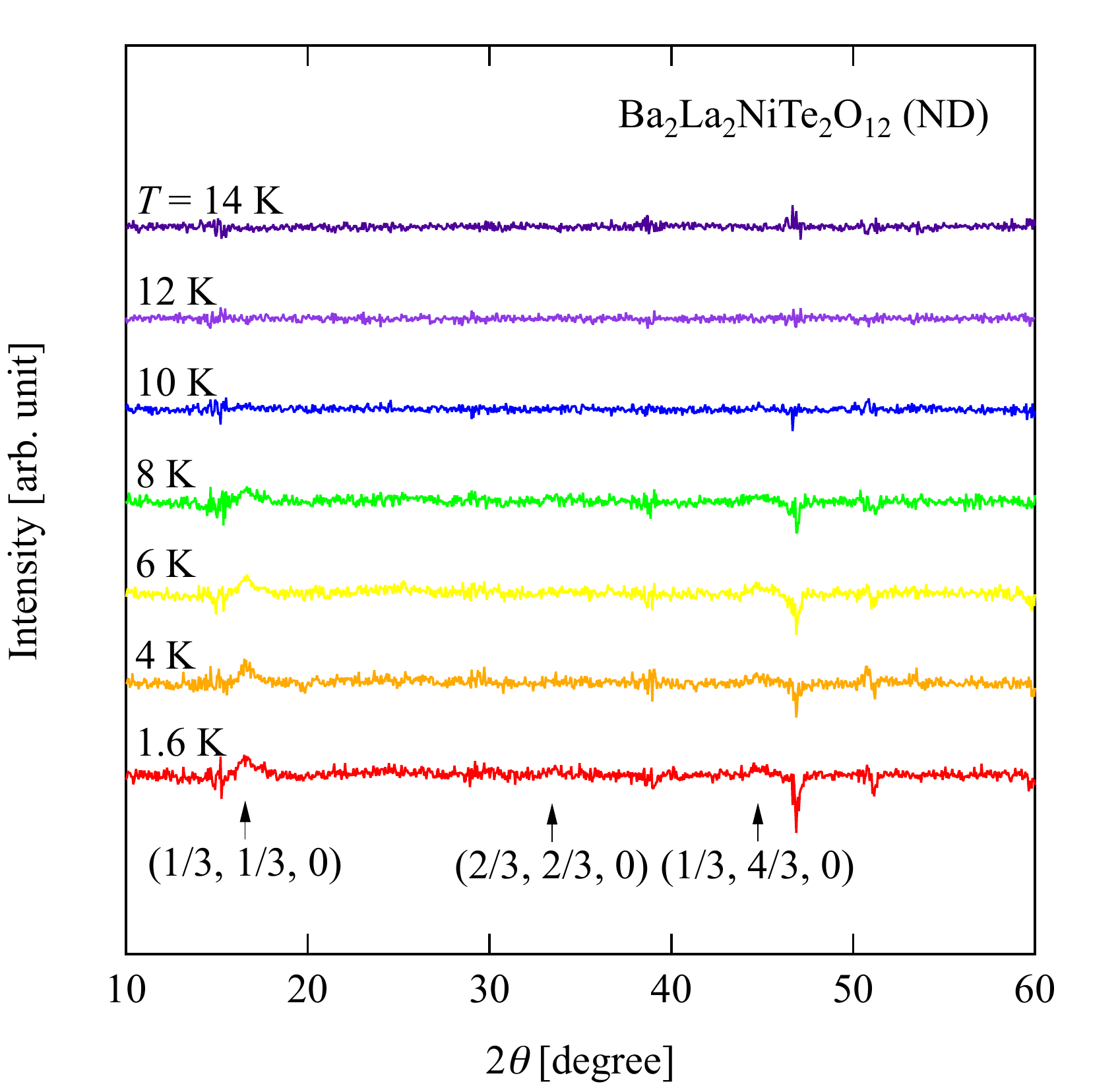}
	\caption{(Color online) ND spectra collected at various temperatures, where the diffraction spectrum for $T\,{>}\,T_{\mathrm{N}1}$ was subtracted as the background. Arrows indicate magnetic Bragg peaks with the indicated wave vectors. Lines for each temperature were arbitrarily shifted in the vertical direction.}
	\label{fig:ND_sa}
\end{figure}

Next, we discuss the magnetic structure in the ordered phases in Ba$_2$La$_2$NiTe$_2$O$_{12}$. The neutron diffraction intensities averaged over $T\,{=}\,14, 12, 10$ K $({>}\,T_{\mathrm{N}1}\,{=}\,9.8$ K) and $T\,{=}\,6, 4, 1.6$ K $({<}\,T_{\mathrm{N}2}\,{=}\,8.9$ K) are shown in Fig. \ref{fig:ND_TN}. There is a small but obvious difference between these ND intensities. Figure \ref{fig:ND_sa} shows powder ND spectra obtained at various temperatures, where the average of the diffraction spectra obtained at $T\,{=}\,14, 12, 10$ K was subtracted as the background. No magnetic peak is observed for $T\,{\geq}\,10$ K. However, new peaks appear below 8 K, which is just below $T_{\mathrm{N}2}\,{=}\,8.9$ K. Thus, these new peaks can be attributed to magnetic Bragg peaks. Diffraction angles for some possible magnetic Bragg reflections, which are estimated from the lattice constants, are also indicated by arrows in Fig. \ref{fig:ND_TN}. The diffraction angles calculated for $\bvec{q}\,{=}\,(1/3, 1/3, 0)$ and its equivalent points coincide with the experimental results. This indicates that Ba$_2$La$_2$NiTe$_2$O$_{12}$ has a triangular spin structure characterized by the propagation vector $\bvec{q}\,{=}\,(1/3, 1/3, 0)$ in the low temperature phase $T_{\mathrm{N}2}$. This propagation vector is in contrast to $\bvec{q}\,{=}\,(1/3, 1/3, 1/2)$ observed for Ba$_2$La$_2$CoTe$_2$O$_{12}$~\cite{Kojima}. The propagation vector $\bvec{q}\,{=}\,(1/3, 1/3, 0)$ observed for Ba$_2$La$_2$NiTe$_2$O$_{12}$ implies that the Y-like triangular structures shown in Fig. \ref{fig:EasyAxis}(a) are ferromagnetically stacked along the $c$ axis; thus, the weak resultant magnetic moments induced in the triangular layers are summed to produce a net moment along the $c$ axis. This spin structure is consistent with the weak magnetic moment observed by magnetization measurement (see Figs. \ref{fig:MT} and \ref{fig:MH}). In addition, we attempted to refine the size of the ordered magnetic moment of Ni$^{2+}$ by the magnetic structure analysis of the ND data but failed owing to the weakness of the magnetic peaks.

\subsection{Density functional theory calculations}

\begin{figure}[t!]
	\centering
	\includegraphics[scale=\scaleoffigband]{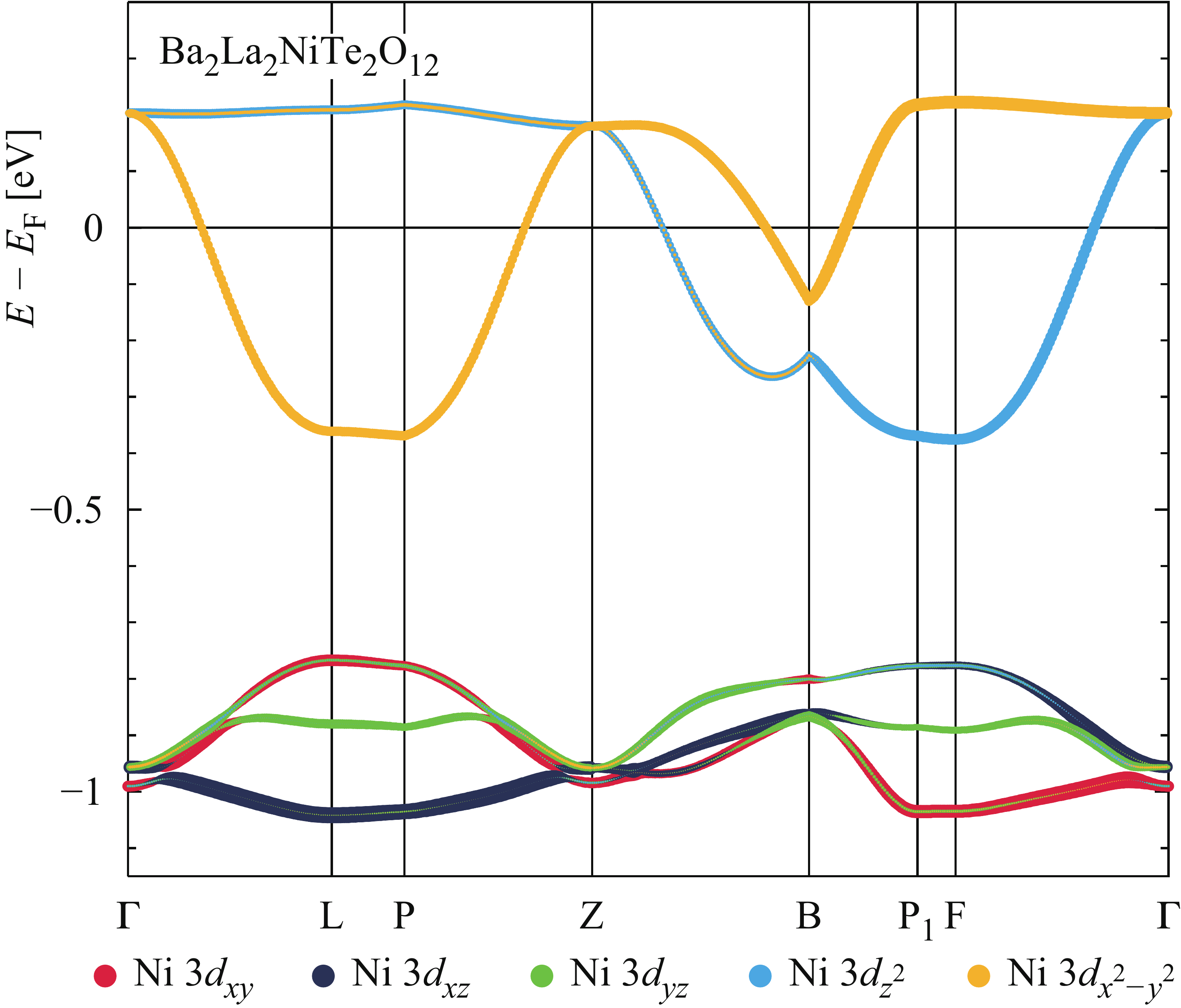}
	\caption{(Color online) Band structure of Ba$_2$La$_2$NiTe$_2$O$_{12}$ obtained from GGA calculations. Orbital weights for Ni $3d$ orbitals are marked. The high-symmetry points for the rhombohedral space group are explained in the text.}
	\label{fig:Bands_BLNTO}
\end{figure}

The band structure of Ba$_2$La$_2$NiTe$_2$O$_{12}$ is shown in Fig.~\ref{fig:Bands_BLNTO}. There are five bands with dominant Ni $3d$ character from the one Ni$^{2+}$ ion in the unit cell. High-symmetry points in the Brillouin zone for the rhombohedral space group $R\,\bar{3}$ are named following Ref.~\cite{Setyawan2010}: $L\,{=}\,(1/2, 0, 0)$, $P\,{=}\,(\eta,\nu,\nu)$, $Z\,{=}\,(1/2,1/2,1/2)$, $B\,{=}\,(\eta,1/2,1-\eta)$, $P_1\,{=}\,(1-\nu,1-\nu,1-\eta)$ and $F\,{=}\,(1/2,1/2,0)$, where $\eta\,{=}\,(1+4\cos\alpha)/(2+4\cos\alpha)$, $\nu\,{=}\,3/4-\eta/2$ and $\alpha\,{=}\,33.889^\circ$ for Ba$_2$La$_2$NiTe$_2$O$_{12}$. Crossing the Fermi level, there are two bands of Ni $e_g$ character, and below there are three bands of Ni $t_{2g}$ character. The width of the two $e_g$ bands is $W=0.6$ eV, three times as large as the band width $W=0.2$ eV in Ba$_2$La$_2$NiW$_2$O$_{12}$ (see Fig. \ref{fig:Bands_BLNWO} in the Appendix). As the hopping parameter, and thus the band width, enters the second-order perturbation estimate of the superexchange quadratically, we can expect the exchange couplings of Ba$_2$La$_2$NiTe$_2$O$_{12}$ to be almost an order of magnitude larger than those of Ba$_2$La$_2$NiW$_2$O$_{12}$.

\begin{figure}[t!]
	\centering
	\includegraphics[scale=\scaleoffig]{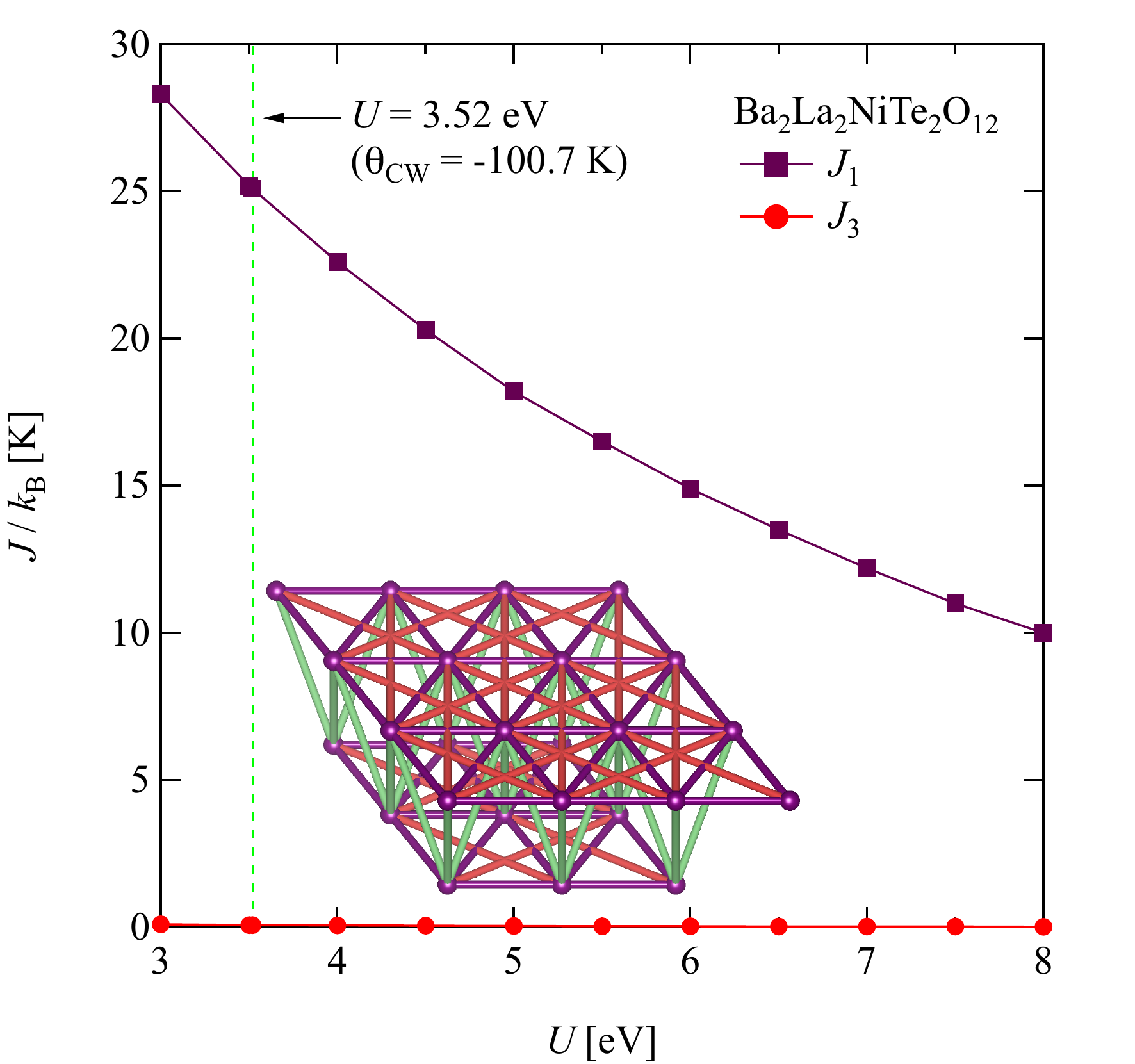}
	\caption{(Color online) In-plane exchange couplings of Ba$_2$La$_2$NiTe$_2$O$_{12}$. The vertical line indicates the $U$ value at which the experimental Curie--Weiss temperature is realized.}
	\label{fig:DFT_J1J3}
\end{figure}

We now proceed to determine the Heisenberg Hamiltonian parameters of Ba$_2$La$_2$NiTe$_2$O$_{12}$ using energy mapping.  We fit all-electron DFT total energies to the Heisenberg Hamiltonian in the form
\begin{equation}
	H=\sum_{i<j} J_{ij} \bvec{S}_i\cdot\bvec{S}_j\,.
\end{equation}
We find that the total moments in all our calculations are exact multiples of $2\mu_{\mathrm{B}}$ as all the nickel moments are exactly $S=1$, and all the fits are very good, resulting in very low statistical errors. We first use a supercell with four Ni$^{2+}$ ions to determine the two in-plane exchange couplings $J_1$ and $J_3$, where we index the couplings with increasing Ni$-$Ni distance. The geometry of the Ni$^{2+}$ ions in Ba$_2$La$_2$NiTe$_2$O$_{12}$ is shown as an inset in Fig. \ref{fig:DFT_J1J3}.

\begin{table}[t!]
	\centering
	\caption{Exchange couplings of Ba$_2$La$_2$NiTe$_2$O$_{12}$, calculated within GGA+U at $J_H=0.88$ eV using a $6\times6\times6$ $k$ mesh in a supercell containing four Ni$^{2+}$ sites. The last row contains the Ni$-$Ni distances, which identify the exchange paths. The errors shown are only the statistical errors arising from the energy mapping.}
	\label{tab:DFT1}
	\begin{tabular}{c|rrr|r}
		\hline\hline
		$U$ [eV] & $J_1/k_{\mathrm{B}}$ [K] & $J_2/k_{\mathrm{B}}$ [K] & $J_3/k_{\mathrm{B}}$ [K] & $\Theta_{\mathrm{CW}}$ [K] \\\hline
		3    & 28.3(1) & - & 0.09(1) & -113 \\
		3.5  & 25.2(1) & - & 0.07(1)  & -101 \\
		{\bf 3.52}  & {\bf 25.1(1)} & - & {\bf 0.07(1)}  & {\bf -100.7} \\
		4    & 22.6(1) & - & 0.06(1) &  -91 \\
		4.5  & 20.3(1) & - & 0.05(1) &  -81 \\
		5    & 18.2(1) & - & 0.04(1) &  -73 \\
		5.5  & 16.5(1) & - & 0.03(1) &  -66 \\
		6    & 14.9(1) & - & 0.03(1) &  -60 \\
		6.5  & 13.5(1) & - & 0.02(1) &  -54 \\
		7    & 12.2(1) & - & 0.02(1) &  -49 \\
		7.5  & 11.0(1) & - & 0.02(1) &  -44 \\
		8    & 10.0(1) & - & 0.01(1) &  -40 \\
		\hline\hline
		$d_{\mathrm{Ni}-\mathrm{Ni}}$ [\AA] & 5.66827 & 9.72442 & 9.81773 & \\
		\hline\hline
	\end{tabular}
\end{table}

The values of the exchange constants are given in Table \ref{tab:DFT1}. The values of $J_i$ are given with respect to spin operators of length $S=1$. Note that if the Hamiltonian is written as $\sum_{ij}$, counting every bond twice, then the values of $J_i$ need to be divided by two. The Curie--Weiss temperatures are estimated from
\begin{equation}
	\Theta_{\mathrm{CW}} = -\frac{2}{3}S(S+1)(3J_1+3J_2+3J_3),
\end{equation}
where $S=1$.

The calculated exchange couplings are shown graphically in Fig. \ref{fig:DFT_J1J3}. The statistical errors are smaller than the symbols. The inset shows the nickel sublattice of the defect perovskite Ba$_2$La$_2$NiTe$_2$O$_{12}$ with bonds indicating the first three exchange pathways. The nearest- and next-nearest-neighbor couplings of the triangular lattice are $J_1$ (purple) and $J_3$ (red), respectively. $J_2$ (turquoise) is the first-interlayer coupling.
$U=3.52$ eV was determined to be the value at which the couplings exactly yield the experimental Curie--Weiss temperature $\Theta_{\mathrm{CW}}=-100.7$ K. 

\begin{table}[t!]
	\centering
	\caption{Exchange couplings of Ba$_2$La$_2$NiTe$_2$O$_{12}$, calculated within GGA+U at $J_H=0.88$ eV with $4\times4\times4k$ points in a supercell containing six Ni$^{2+}$ sites. The last row contains the Ni$-$Ni distances, which identify the exchange paths. The errors shown are only the statistical errors arising from the energy mapping.}
	\label{tab:DFT2}
	\begin{tabular}{c|rrr|r}
		\hline\hline
		$U$ [eV] & $J_1/k_{\mathrm{B}}$ [K] & $J_2/k_{\mathrm{B}}$ [K] & $J_3/k_{\mathrm{B}}$ [K] & $\Theta_{\mathrm{CW}}$ [K] \\\hline
		3    & 28.25(1) & 0.024(1) & 0.078(1) & -113 \\
		3.5  & 25.21(1) & 0.022(1) & 0.062(1) & -101 \\
		\textbf{3.52} & \textbf{25.09(1)} & \textbf{0.021(1)} & \textbf{0.062(1)} & \textbf{-100.7} \\
		4    & 22.57(1) & 0.018(1) & 0.051(1) &  -91 \\
		4.5  & 20.26(1) & 0.016(1) & 0.039(1) &  -81 \\
		5    & 18.23(1) & 0.014(1) & 0.034(1) &  -73 \\
		\hline\hline
		$d_{\mathrm{Ni}-\mathrm{Ni}}$ [\AA] & 5.66827 & 9.72442 & 9.81773 & \\
		\hline\hline
	\end{tabular}
\end{table}

A larger supercell containing six inequivalent Ni$^{2+}$ sites also allows the determination of the interlayer coupling $J_2$. The result of this calculation is shown in Table \ref{tab:DFT2}. The interlayer coupling turns out to be even smaller than the next-nearest-neighbor coupling $J_3$ in the triangular lattice. However, consistent with the fact that the calculation with the four-Ni$^{2+}$ unit cell does not allow the separation of $J_1$ and $J_2$, meaning that the $J_1$ values in Table \ref{tab:DFT1} actually represent the sum $J_1+J_2$, the new $J_1$ values in Table \ref{tab:DFT2} are very slightly smaller than those in Table \ref{tab:DFT1}. However, this more precise calculation still has not yielded a substantial subleading coupling to the antiferromagnetic $J_1$. The $U$ value that can reproduce the experimental Curie--Weiss temperature $\Theta_{\mathrm{CW}}=-100.7$ K is still $U=3.52$ eV.

From these DFT calculations, Ba$_2$La$_2$NiTe$_2$O$_{12}$ was found to be a pure triangular lattice antiferromagnet with a nearly negligible next-neighbor coupling in the plane. However, as the interlayer Ni-Ni distance is comparable to the in-plane next-neighbor distance, we also determined this additional coupling using larger supercells for the energy mapping. However, these calculations indicate that Ba$_2$La$_2$NiTe$_2$O$_{12}$ is, as a very good approximation, a 2D triangular lattice antiferromagnet. The only interlayer coupling we were able to resolve, $J_2$, is tiny and antiferromagnetic. Thus, the small ferromagnetic coupling between the layers that was experimentally inferred from the weak magnetic moment at zero field could, for example, arise from the as yet unknown $J_4$ at a distance of $d_{\mathrm{Ni-Ni}}=11.256$ \AA.

\begin{figure}[t!]
	\centering
	\includegraphics[scale=\scaleoffigband]{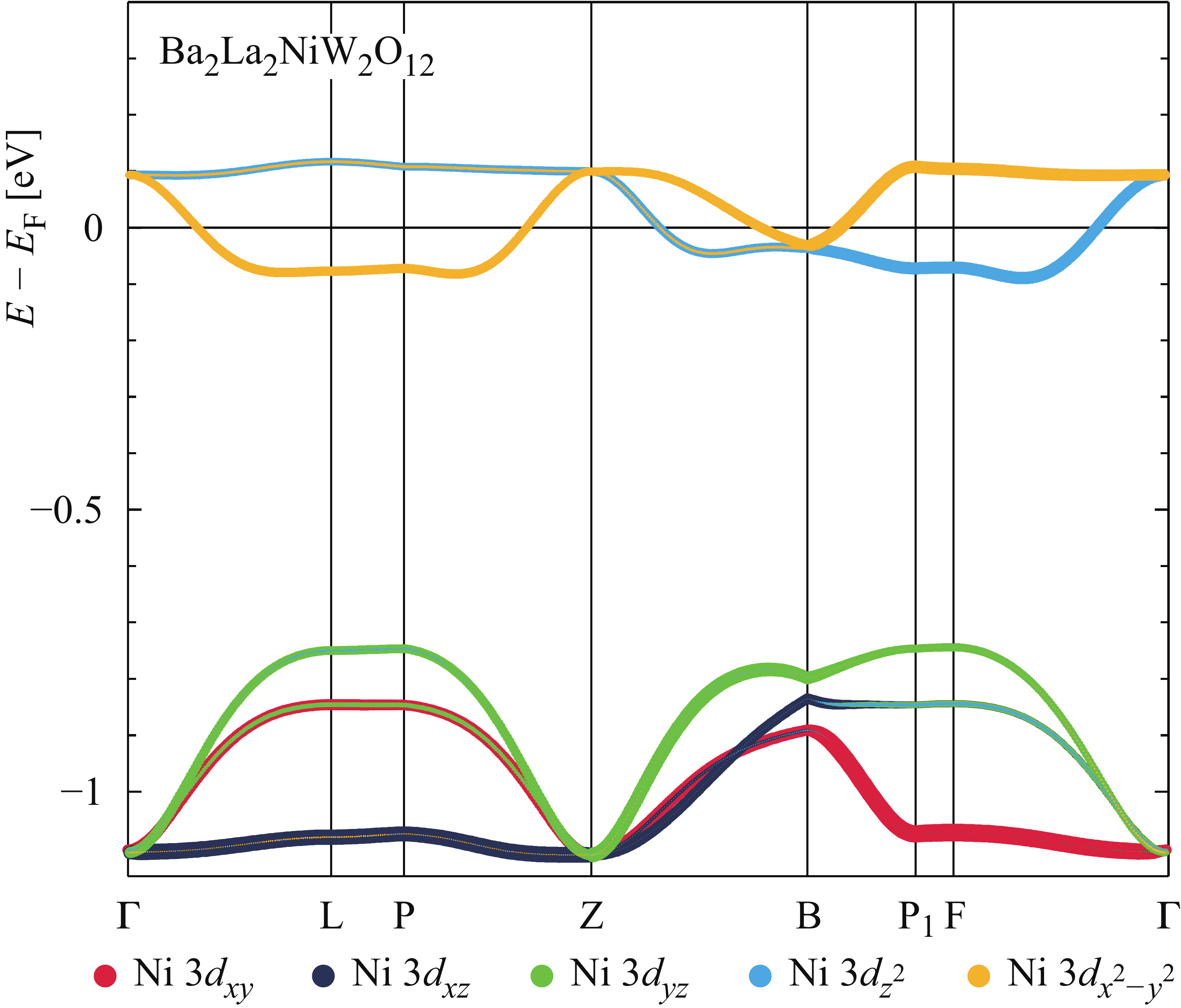}
	\caption{(Color online) Band structure of Ba$_2$La$_2$NiW$_2$O$_{12}$ obtained from GGA calculations. Orbital weights for Ni $3d$ orbitals are marked.}
	\label{fig:Bands_BLNWO}
\end{figure}

\section{Conclusion}
We have reported on the crystal structure and magnetic properties of the spin-1 TLAF Ba$_2$La$_2$NiTe$_2$O$_{12}$ composed of a uniform triangular lattice of Ni$^{2+}$ ions.
We refined the crystal structure parameters by Rietveld analysis using XRD and ND data obtained from a powdered sample. The space group was determined to be $R\bar{3}$. The large negative Weiss constant ${\Theta}_{\mathrm{CW}}\,{\simeq}\,-100$ K for the magnetic susceptibility shows that the predominant exchange interaction is antiferromagnetic and strong, in contrast to Ba$_2$La$_2$NiW$_2$O$_{12}$~\cite{Rawl,Doi}. Specific heat measurement demonstrated that Ba$_2$La$_2$NiTe$_2$O$_{12}$ undergoes successive magnetic phase transitions at $T_{\mathrm{N}1}\,{=}\,9.8$\,K and at $T_{\mathrm{N}2}\,{=}\,8.9$\,K, which arise from the competition between the antiferromagnetic exchange interaction and the single-ion anisotropy of the easy-axis type. From the weak net magnetic moment of $\overline{\Delta M}\,{=}\,0.015\,{\mu_{\mathrm{B}}}/{\mathrm{Ni}}^{2+}$ observed at $T\,{=}\,1.8$ K (${\ll}\,T_{\mathrm{N}2}$), the ratio of single-ion anisotropy to the exchange interaction was estimated as $|D|/J\,{\simeq}\,0.108$. It was found from high-magnetic-field magnetization measurement up to 60 T that the magnetization curve exhibits a wide plateau at one-third of the saturation magnetization, which is characteristic of 2D Heisenberg-like TLAFs. We estimated the exchange interaction $J$ and the $g$-factor as $J/k_{\mathrm{B}}\,{\simeq}\,19$ K and $g\,{\simeq}\,2.3$, respectively, by fitting the theoretical magnetization curve to the experimental data. From the ND measurements at zero magnetic field, the propagation vector in the low-temperature phase for $T<T_{\mathrm{N}2}$ was found to be $\bvec{q}\,{=}\,(1/3, 1/3, 0)$. This result, together with the magnetization and specific heat results, indicates that below $T_{\mathrm{N}2}$, spins form a triangular structure in a plane including the $c$ axis in each triangular layer and these triangular spin structures are ferromagnetically stacked along the $c$ axis.
The DFT calculations demonstrated that the nearest-neighbor exchange interaction is predominant and that the next-nearest-neighbor exchange interaction in the triangular layer and the interlayer exchange interactions are negligible.

\section*{Acknowledgments}
We thank the authors of Ref.~\cite{Richter} for allowing us to use their theoretical calculations of the magnetization process. This work was supported by Grants-in-Aid for Scientific Research (A) (No.~17H01142) and (C) (No.~16K05414) from Japan Society for the Promotion of Science.

\appendix

\section{Electronic structure of Ba$_2$La$_2$NiW$_2$O$_{12}$}
For comparison with the new material Ba$_2$La$_2$NiTe$_2$O$_{12}$, we have determined the electronic structure of Ba$_2$La$_2$NiW$_2$O$_{12}$ using the crystal structure provided in Ref.~\cite{Rawl}. Figure \ref{fig:Bands_BLNWO} shows the bands calculated with the GGA exchange correlation functional. The path through the Brillouin zone is explained in the main text. As in isostructural Ba$_2$La$_2$NiTe$_2$O$_{12}$, two Ni $3d$ bands of $e_g$ character cross the Fermi level. However, the band width is only 0.2 eV, indicating rather small effective hopping parameters between Ni $e_g$ orbitals compared to  Ba$_2$La$_2$NiTe$_2$O$_{12}$.

\end{document}